\documentclass{aa}
\usepackage{graphicx}
\usepackage{txfonts}
\usepackage{tabularx}
\newcommand{\Msun}{\ensuremath{\mathrm{M}_\odot}}

\newcommand{\lsim}{\mathrel{\hbox{\rlap{\lower.55ex \hbox {$\sim$}}
 \kern-.3em \raise.4ex \hbox{$<$}}}}
\newcommand{\gsim}{\mathrel{\hbox{\rlap{\lower.55ex \hbox {$\sim$}}
 \kern-.3em \raise.4ex \hbox{$>$}}}}

\begin{document}
   \title{On the evolution of rapidly rotating massive white dwarfs %
      towards supernovae or collapses}


   \author{S.-C. Yoon
         \inst{1, 2}
          \and
          N. Langer
          \inst{1}
          }

   \offprints{S.-C. Yoon}

   \institute{Astronomical Institute, Utrecht University,
              Princetonplein 5, NL-3584 CC, Utrecht, The Netherlands
         \and
              \emph{present address:} Astronomical Institute "Anton Pannekoek", University of Amsterdam, 
               Kruislaan 403, 1098 SJ, Amsterdam, The Netherlands 
              \\
              \email{scyoon@science.uva.nl; n.langer@astro.uu.nl}
             }

   \date{Received ; accepted }

   \abstract{
     A recent study by Yoon \& Langer (\cite{Yoon04a}) indicated that 
     the inner cores of rapidly accreting ($\dot M > 10^{-7}~\mathrm{M_\odot~yr^{-1}}$) CO white dwarfs
     may rotate differentially, with a shear rate near the threshold value for the onset
     of the dynamical shear instability. Such differentially rotating white dwarfs obtain
     critical masses for thermonuclear explosion or electron-capture induced collapse
     which significantly exceed the canonical Chandrasekhar limit.
     Here, we construct two-dimensional differentially rotating white dwarf models
     with rotation laws resembling those of the one-dimensional models of Yoon \& Langer~(\cite{Yoon04a}).
     We derive analytic relations between the white dwarf mass, its angular 
     momentum, and its rotational-, gravitational- and binding energy. 
     We show that these relations are applicable for a wide range of 
     angular velocity profiles, including solid body rotation.
     Taken at a central density of $2\times10^9~\mathrm{g~cm^{-3}}$
     they specify initial models for the thermonuclear explosion of rotating CO white dwarfs.
     At $\rho_\mathrm{c} = 10^{10}~\mathrm{g~cm^{-3}}$ and $4\times10^{9}~\mathrm{g~cm^{-3}}$, 
     they give criteria for the electron-capture induced collapse of 
     rotating CO and ONeMg white dwarfs, respectively. 
     We demonstrate that pre-explosion and pre-collapse conditions of both, rigidly and
     differentially rotating white dwarfs
     are well established by the present work, which may facilitate future
     multi-dimensional simulations of Type~Ia supernova explosions and
     studies of the formation of millisecond pulsars and gamma-ray bursts from collapsing
     white dwarfs.
     Our results lead us to suggest various possible evolutionary 
     scenarios for progenitors of Type~Ia supernovae, leading to a new paradigm
     of a variable mass of exploding white dwarfs, at values well above the classical
     Chandrasekhar mass. Based on our 2D-models, we argue for the
     supernova peak brightness being proportional to the white dwarf mass, which could
     explain various aspects of the diversity of Type~Ia supernovae, such as their
     variation in brightness, the dependence of their mean luminosity on
     the host galaxy type, and the weak correlation between ejecta velocity and
     peak brightness.
   }
   \titlerunning{Rapidly rotating massive white dwarfs}
   \authorrunning{Yoon \& Langer}
   \maketitle
%

\section{Introduction}

Rotation, a universal phenomenon in celestial objects, 
is one of the most classical topics in stellar astrophysics (Tassoul~\cite{Tassoul78}).
Today, rotation is considered as an important ingredient to stellar evolution
(Heger \& Langer~\cite{Heger00}; Maeder \& Meynet~\cite{Maeder00}).
Key aspects of massive star deaths -- notably the production
of millisecond pulsars, hypernovae and gamma-ray bursts --
require rapid rotation
(e.g., Woosley~\cite{Woosley93}; Paczy\'nski~\cite{Paczynski98}; Hirschi et al.~\cite{Hirschi04}; 
Heger et al.~\cite{Heger04}; Petrovic et al.~\cite{Petrovic05b}).
In close binary systems, effects of rotation become even more
important, due to the exchange of mass and angular momentum
between the stellar components (Langer et al.~\cite{Langer04}).
For instance, the accreting stars in massive close binaries may be spun up to 
critical rotation, which influences significantly their evolution
and final fate (Petrovic et al.~\cite{Petrovic05a}). 

It is hence surprising that for the last decades, the effects of rotation
on Type~Ia supernova (SN~Ia) progenitors --- objects crucial
for understanding chemical evolution as well as the expansion history of the universe 
(e.g. Leibundgut~\cite{Leibundgut01}) ---
have not been considered in detail. 
The currently favored Single Degenerate scenario for the evolution of SNe~Ia progenitors  
assumes that the white dwarf accretes matter 
from a non-degenerate companion with high 
accretion rates ($\dot{M} = 10^{-7} \dots  10^{-6}~\mathrm{M_\odot}$), to
grow to the Chandrasekhar limit by steady nuclear shell burning
(Hachisu et al.~\cite{Hachisu96}; Li \& van den Heuvel~\cite{Li97};
Langer et al.~\cite{Langer00}; Yoon \& Langer~\cite{Yoon03}; Han \& Podsiadlowski~\cite{Han04}).  
The white dwarf is supposed to accrete matter through
a Keplerian disk, implying that the accreted matter
carries large amounts of angular momentum
(e.g. Langer et al.~\cite{Langer00}). Here, as well as
in alternative SN~Ia progenitor scenarios, such as the sub-Chandrasekhar Mass scenario 
or the Double Degenerate scenario
(see below), a spin-up of the accreting white dwarfs seems unavoidable.
As an understanding of their diversity 
is essential for the use of SNe~Ia to measure cosmological distances
(e.g. Leibundgut~\cite{Leibundgut01}), 
the effects of rotation need to be carefully studied, since they could
influence both, the SNe~Ia progenitor evolution
and the thermonuclear explosion of the white dwarf.
 
During the last two years, various systematic investigations
of the effects of the white dwarf spin-up due to accretion 
in the context of SNe~Ia progenitor evolution have emerged
(Piersanti~\cite{Piersanti03a}, \cite{Piersanti03b}; Uenishi~\cite{Uenishi03}; 
Yoon \& Langer~\cite{Yoon04a}, \cite{Yoon04b}; Yoon et al.~\cite{Yoon04c};
Saio \& Nomoto~\cite{Saio04}). 
These results may demand surprising modifications of the standard picture of SNe~Ia progenitors. 

Yoon, Langer \& Scheithauer~(\cite{Yoon04c}) showed that the
shell burning in rapidly accreting white dwarfs
is stabilized by the accretion induced spin-up, as
the shell source is widened through
rotationally induced chemical mixing and decompressed by 
the centrifugal support (cf. Yoon, Langer \& van der Sluys~\cite{Yoon04d}).
This result supports the key assumption of the Single Degenerate 
scenario that white dwarfs can efficiently grow in mass by shell burning,
which has been challenged by the finding of generally 
unstable helium shell burning and subsequent mass loss
in non-rotating white dwarfs
(Cassisi et al.~\cite{Cassisi98}; Kato \& Hachisu~\cite{Kato04}).

In the sub-Chandrasekhar Mass scenario, the detonation of a thick helium layer
triggers the explosion of a sub-Chandrasekhar mass CO core
(e.g. Woosley \& Weaver~\cite{Woosley94}).
Here, Yoon \& Langer~(\cite{Yoon04b}) showed that
as the white dwarf is spun up by accretion at the moderate
rates required to build up the degenerate helium layer, the latter is
efficiently heated by the dissipation of rotational energy.
Consequently, helium ignites much earlier and under less degenerate conditions
than in the corresponding non-rotating case, which 
avoids a supernova explosion altogether.

Most intriguingly, a study by Yoon \& Langer~(\cite{Yoon04a}; hereafter Paper~I)
indicated that accreting white dwarfs with $\dot{M} \gsim 10^{-7}~\mathrm{M_\odot~yr^{-1}}$
may rotate differentially throughout. 
They found that, although the strongly degenerate inner core is susceptible
to the dynamical shear instability (DSI), other instabilities such as 
Eddington-Sweet circulations, the secular shear instability 
and the Goldreich, Schubert and Fricke instability
are relevant for angular momentum redistribution at the required time scales ($\sim 10^6~\mathrm{yr}$)
only in the non-degenerate outer layers.
Accordingly, the shear rate in the inner core of the white dwarf corresponds to the shear rate near the 
threshold value for the onset of the DSI (Paper~I).
Since differentially rotating white dwarf
can be dynamically stable even up to $4.0~\mathrm{M_\odot}$, the 
masses of exploding white dwarfs could be significantly larger than
the canonical Chandrasekhar limit of $\sim$1.4~\Msun (cf. Paper~I)..
This may require to supersede the long-standing paradigm 
of Chandrasekhar-mass white dwarfs as SN~Ia progenitors
with a new paradigm of SN~Ia progenitors as white dwarfs with a variety of masses
at super-Chandrasekhar mass values.

The possibility of super-Chandrasekhar mass white dwarfs is not new;
already in the 1960ies, e.g. Ostriker \& Bodenheimer~(\cite{Ostriker68a})
constructed differentially rotating 2-D white dwarf models of up to 4.1\Msun.
However, the internal angular momentum distribution in these models had to be
assumed and was chosen such that the spin rate decreases outward ---
conversely to the result of evolutionary considerations (Paper~I).
More recently, Uenishi et al.~(\cite{Uenishi03}) provided
2-D white dwarf models, assuming a rotation law which
introduces a discontinuity between the non-rotating inner core
and the rigidly rotating spun-up envelope. However, due to this
assumption, the maximum possible white dwarf mass became 1.47\Msun,
which strongly restricts the diversity of possible explosion masses.
In the present paper, we construct 2-D white dwarf models with
rotation laws that mimic the 1-D results of Paper~I and thus follow
from calculating the evolution of the internal white dwarf angular momentum profile
throughout the accretion phase. 

The remainder of this paper is organized as follows.
In Sect.~\ref{sect:constructing_2-D_models}, 
we explain the numerical method
and physical assumptions for constructing 2-D white dwarf models.
Basic model properties are described in Sect.~\ref{sect:results},
and, in the next section (Sect.~\ref{sect:relations}), 
we present the quantitative criteria
for thermonuclear explosion and electron-capture (EC) induced collapse of white dwarfs
in terms of the white dwarf mass and total angular momentum.
In Sect.~\ref{sect:comparison}, the 1-D models of Paper~I are compared
to the 2-D models of the present study.
We discuss implications of our results
for the evolution of accreting white dwarfs, 
and suggest new evolutionary scenarios for SNe~Ia progenitors  
in Sect~\ref{sect:implication_SNIa}. Possible consequences for the diversity of 
SN~Ia are elaborated in Sect.~\ref{sect:diversity}, while implications
of the EC induced collapse of differentially rotating
white dwarfs are briefly discussed in Sect.~\ref{sect:collapse}.
We summarize our main conclusions in Sect.~\ref{sect:conclusion}.

\section{Constructing 2-D models}\label{sect:constructing_2-D_models}

\subsection{Basic assumptions}\label{sect:basic_assumtions}

Rotating stars are called \emph{barotropic} if the following two conditions 
are satisfied (Tassoul~\cite{Tassoul78}): 
\begin{equation}\label{eq1}
P=P(\rho) ~~~ \mathrm{and} ~~~ \frac{\partial \Omega}{\partial z} = 0, 
\end{equation}
where 
$P$ is the pressure, $\rho$ the density
$\Omega$ the angular velocity and $z$ the distance to the equatorial plane, respectively. 
Degenerate cores of accreting white dwarfs are thought to be barotropic 
on two accounts. First, the equation of state of highly electron-degenerate matter is given 
by $P=P(\rho)$.  Second, perturbations which create 
$\frac{\partial \Omega}{\partial z} \ne 0$ will decay on a short time scale in degenerate matter,
as discussed in Kippenhahn \& M\"ollenhof~(\cite{Kippenhahn74}) and Durisen~(\cite{Durisen77}). 
Since the mass of the non-degenerate outer envelope in an accreting white dwarf 
is usually small (Paper~I; Yoon et al.~\cite{Yoon04c}), 
the barotropic condition may
be a good approximation for describing the whole white dwarf.
For constructing axisymmetric 2-D white dwarf models,
we adopt cylindrical rotation as described below, 
and the equation of state of a completely degenerate electron
gas, as in Ostriker \& Bodenheimer (\cite{Ostriker68a}):
\begin{equation}\label{eq2}
P = A[x(2x^2-3)(x^2+1)^{1/2}+3\sinh^{-1}x]\,;\, ~ \rho = Bx^3~, 
\end{equation}
where $A = 6.01\times10^{22}~\mathrm{dynes~cm^{-2}}$ and 
$B=9.82\times10^5\mu_\mathrm{e}~\mathrm{g~cm^{-3}}$. Here
we adopt $\mu_\mathrm{e} = 2$ for all calculations.

\subsection{The rotation law}\label{sect:rotation_law}
According to Paper~I, the rotation law in a white dwarf which accretes matter and angular momentum
with accretion rates higher than $10^{-7}~\mathrm{M}_\odot/\mathrm{yr}$
is critically affected by the angular momentum transport due to the dynamical shear instability (DSI) 
in the inner core, and
due to the secular shear instability, Eddington-Sweet circulations and the GSF instability in 
the outer layers. In particular, the shear rate in the inner core 
remains near the threshold value for the onset of the DSI ($:= \sigma_\mathrm{DSI,crit}$) 
during the accretion phase (see Paper~I for more details). Another important feature
of the rotation law
is the presence of an absolute maximum in the angular velocity profile 
just above the shear-unstable core, which occurs
due to the compression of the accreted layers with time (cf. also Durisen~(\cite{Durisen77})
and see Fig. 9 in Paper~I).
From these constraints, we can construct a realistic rotation law as follows. 

We define $s_\mathrm{p}$ as the position of the maximum angular velocity.
As discussed in Paper~I, above this position the 
buoyancy force is strong enough to assure stability against the DSI.
In the white dwarf models in Paper~I, 
this position is linked to layers with a density as low as 
several percents of the central density ($\rho_\mathrm{c}$).
Therefore, we determine $s_\mathrm{p}$ such that 
\begin{equation}\label{eq3} 
\rho(s=s_\mathrm{p}, z=0) = f_\mathrm{p}\,\rho_\mathrm{c}~.
\end{equation}
In our models, we use $f_\mathrm{p} = 0.1$ in most cases.
Then, $\Omega(s)$ is determined 
in the range of $0 \leq s \leq s_\mathrm{p}$ from 
the fact that,
in the inner core, the shear rate is given by
\begin{equation}\label{eq4}
\frac{d\Omega(s)}{d\ln s} = f_\mathrm{sh}\sigma_\mathrm{DSI, crit}~.
\end{equation}
Here, $s$ is the distance from the rotation axis in the cylindrical coordinate and
the dimensionless parameter $f_\mathrm{sh}$ ($\leq 1.0$) describes the deviation 
of the shear rate from  $\sigma_\mathrm{DSI, crit}$.
We use $f_\mathrm{sh} = 1.0$ except for some test cases. 
The threshold value of the shear rate for the onset of
the DSI is given as $\sigma_\mathrm{DSI, crit} = \sqrt{N^2/R_\mathrm{i,c}}$, 
where $N^2$ is the Brunt-V$\ddot{\rm a}$is$\ddot{\rm a}$l$\ddot{\rm a}$ frequency and
$R_\mathrm{i,c}$ the critical Richardson number (see Paper~I for more details). 
From this we get
\begin{equation}\label{eq5}
\Omega(s) =  \Omega_\mathrm{c} + \int_{0}^{s} \frac{f_\mathrm{sh}\sigma_\mathrm{DSI, crit}}{s'} ds' ~, 
\end{equation}
where $\Omega_\mathrm{c}$ is the angular velocity at the center. 
In order to ensure that $d\Omega/ds = 0$ at $s=s_\mathrm{p}$, which is the case
in 1-D models of Paper~I, 
we change the values of $\sigma_\mathrm{DSI, crit}$ around $s_\mathrm{p}$
such that  
\begin{equation}\label{eq6}
\sigma_\mathrm{DSI, crit}(s) = \left\{ \begin{array}{ll} 
\sqrt{N^2/R_\mathrm{i,c}}  & \textrm{if}~~ 0 \leq s \leq s_\mathrm{b} \\
\sqrt{N^2/R_\mathrm{i,c}}(s-s_\mathrm{p})/(s_\mathrm{b}-s_\mathrm{p}) 
& \textrm{if}~~ s_\mathrm{b} < s \leq s_\mathrm{p} \end{array} \right.~, 
\end{equation}
where $s_\mathrm{b}$ is a certain location near to $s_\mathrm{p}$. 
In this study, we define $s_\mathrm{b}$ as the fifth neighboring grid point
from $s_\mathrm{p}$ (while 149 grid points in $s$-direction are used).
The calculation of $N^2$ is carried out by the use of the equation of state
given by Blinnikov et al.~(\cite{Blinnikov96}), using the equatorial values of $\rho$ 
and assuming $T=10^8$~K which is 
typical in the inner core of the accreting white dwarf models in Paper I.
For the critical Richardson number, $R_\mathrm{i,c} = 0.25$ is used.
The integration in Eq.~(\ref{eq5}) is carried out numerically 
by fitting $f_\mathrm{sh}\sigma_\mathrm{DSI, crit}/s$ 
with the Chebyshev polynomial (cf. Press et al.~\cite{Press86}). 

At the surface
the white dwarf equator is assumed to rotate at a certain fraction ($f_\mathrm{K}$) 
of the Keplerian value, such that
\begin{equation}\label{eq7}
\Omega(R_\mathrm{e}) = f_{\rm K} \Omega_\mathrm{K}(R_\mathrm{e})~, 
\end{equation}
where $R_\mathrm{e}$ is the equatorial radius of the white dwarf, and $\Omega_\mathrm{K}(R_\mathrm{e})
= \sqrt{GM/R_\mathrm{e}^3}$. 
An accreting white dwarf may be expected to rotate near the critical velocity at its surface, since
accretion through a Kelperian disk will carry angular momentum with the local Keplerian value.
Therefore, we use $f_\mathrm{K} = 0.95$ in most cases.

It remains to determine $\Omega(s)$ between $s_{\rm p}$ and $R_\mathrm{e}$. 
We impose an analytical form to $\Omega(s)$ in this region as follows: 
\begin{equation}\label{eq8}
\Omega(s)/\Omega_\mathrm{K}(s) = \Omega(s_\mathrm{p})/\Omega_\mathrm{K}(s_\mathrm{P}) + C(s-s_\mathrm{p})^a~, 
\end{equation}
where $\Omega_\mathrm{K}(s)$ is the local Keplerian value. 
The constant $C$ can be determined with a given value of $a$ as
\begin{equation}\label{eq9}
C = \frac{f_\mathrm{K} - \Omega(s_\mathrm{p})/\Omega_\mathrm{K}(s_\mathrm{p})} 
         {(R_\mathrm{e} - s_\mathrm{p})^a}~.
\end{equation}
The white dwarf structure is not significantly affected by
the choice of the index $a$, because of the constraints imposed 
by $\Omega(s_\mathrm{p})$ and  $\Omega(R_\mathrm{e})$ at each edge. 
Furthermore, the
density in the outer region ($s>s_\mathrm{p}$) is lower than the inner core, 
and the role of the rotation law of this region for the white dwarf structure is less important 
than that of the inner core.  
In the present study, $a=1.2$ is usually employed.

In summary, we determine the rotation law as follows.
In the inner region ($ 0 \leq s \leq s_\mathrm{p}$), $\Omega(s)$
is determined by the dynamical shear instability (
Eqs.~(\ref{eq5}) and~(\ref{eq6}))  for a given $\Omega_\mathrm{c}$. 
In the outer region ($ s_\mathrm{p} \leq s \leq R_\mathrm{e}$), 
$\Omega(s)$ is determined by Eq.~(\ref{eq8}) where  $C$ 
is given by Eq.~(\ref{eq9}).

\subsection{SCF method}\label{sect:SCF_method}
For constructing 2-D white dwarf models with the prescribed rotation law, 
we use the so-called self-consistent field (SCF) method  which has been widely employed
for modeling rotating stars
(Ostriker \& Mark~\cite{Ostriker68b}; Hachisu~\cite{Hachisu86}). 
This method can be most easily employed with barotropic stars, 
which satisfy the relation $\rho = \rho(\Psi)$, 
where the total potential
$\Psi$ is the sum of the gravitational potential  $\Phi$
and the centrifugal potential $\Theta$:
\begin{equation}\label{eq10}
\Psi(\mu,r) = \Phi(\mu,r) + \Theta(s)~,
\end{equation}
where
\begin{equation}\label{eq11}
\Phi(\mu,r) = G\int\frac{\rho(\mathbf{r'})}{|\mathbf{r}-\mathbf{r'}|} d^3\mathbf{r'}
\end{equation}
and
\begin{equation}\label{eq12}
\Theta(s) = \int_0^s \Omega(s')^2s' ds'~.
\end{equation}
Here, $\mu$($:=\cos\theta$) and $r$ represent the spherical coordinates. 
The distance from the rotation axis $s$ is given by $s = r(1-\mu^2)^{1/2}$. 
In our study, the gravitational potential $\Phi$ is calculated
using Simpson's formula following Hachisu~(\cite{Hachisu86}).
For the centrifugal potential $\Theta$, the integration in Eq.~(\ref{eq12})
is carried out by approximating the integrand by 
Chebyshev polynomials.

The SCF method then proceeds as follows. 
Once we have $\Psi$ from 
a given trial density distribution $\rho(\mu,r)$ 
and a given angular velocity profile, 
a better value for the density $\rho$ can be given analytically as a function of $\Psi$
(see Ostriker \& Bodenheimer~\cite{Ostriker68a}).
This new density distribution in turn gives a new value of $\Psi$.
This procedure is iterated until a good convergence level is achieved.
In the present study, the iteration continues until the following conditions
are satisfied:
\begin{equation}\label{eq13}
\left|\frac{\Psi^\mathrm{new}-\Psi^\mathrm{old}}{\Psi^\mathrm{new}} \right|_\mathrm{max} < 5\cdot10^{-5},~\mathrm{and}~ 
\left|\frac{C - <C>}{<C>}\right|_\mathrm{max} < 5\cdot10^{-5},
\end{equation}
where $C$, which should be constant in ideal equilibrium structures, 
is defined as 
\begin{equation}\label{eq14}
C = \int \rho^{-1} dP - \Phi - \Theta~, 
\end{equation}
and $<C>$ denotes its mean value.
One additional condition is imposed by the virial relation such that
\begin{equation}\label{eq15}
\frac{|2T-W+3\Pi|}{W} < 10^{-3}~, 
\end{equation}
where $T$ and $W$ denote rotational- and gravitational energy, respectively, 
and $\Pi$ the volume integral of the pressure.
Note that here we define gravitational energy as a positive value for convenience.
Hereafter, we will refer to the above explained procedure 
as \emph{SCF iteration}.

\subsection{Further details}\label{sect:further_details}

We use the maximum density ($\rho_\mathrm{max}$), the gravitational constant ($G$)
and the equatorial radius at the surface of the white dwarf ($R_\mathrm{e}$)
for the basis of dimensional variables for numerical calculations, as in Hachisu~(\cite{Hachisu86}). 
With this choice, the physical variables can be expressed in a dimensionless form 
such that $\hat{\rho} = \rho/\rho_\mathrm{max}$, $\hat{s} = s/R_\mathrm{e}$, 
$\hat{j} = j/\sqrt{G\rho_\mathrm{max}}R_\mathrm{e}^2$, 
$\hat{\Omega} = \Omega/\sqrt{G\rho_\mathrm{max}}$, $\hat{\Psi} =\Psi/GR_\mathrm{e}^2\rho_\mathrm{max}$, 
and so on. As pointed out by Hachisu (\cite{Hachisu86}), 
using a fixed value of $\rho_\mathrm{max}$ 
instead of the total mass $M$ makes the numerical convergence faster, especially 
because the equatorial radius
$R_\mathrm{e}$ can be given analytically as
\begin{equation}\label{eq16}
R_\mathrm{e} = \left\{\frac{8A}{B}
                \frac{[1+(\rho_\mathrm{max})^{2/3}]^{1/2} -1}
                     {G\rho_\mathrm{max}(\hat{\Psi}_\mathrm{max} - \hat{\Psi}_\mathrm{s})}
                \right\}^{1/2}~,  
\end{equation}
where $\hat{\Psi}_\mathrm{s}$ indicates the boundary value.

Hachisu also chose the axis ratio as an initially fixed parameter for determining the equilibrium structure. 
This approach is possible only when the rotation law is given analytically, 
and can not be applied for the rotation law defined 
in Sect.~\ref{sect:rotation_law}.
Instead, we choose the central angular velocity $\Omega_\mathrm{c}$
as an initial parameter. 
Therefore, in our models, 
initially fixed maximum density $\rho_\mathrm{max}$
and central angular velocity $\Omega_\mathrm{c}$ 
determine  the equilibrium structure of the white dwarf, together
with given values of $f_\mathrm{p}$ (Eq.~\ref{eq3}), 
$f_\mathrm{sh}$ (Eq.~\ref{eq4}), $f_{\rm K}$ (Eq.~\ref{eq7})
and $a$ (Eq.~\ref{eq9}). 
More specifically, at a given $\rho_\mathrm{max}$, the total angular momentum $J$
and the white dwarf mass $M$ increase with increasing $\Omega_\mathrm{c}$.

It is found that, 
when the centrifugal potential 
$\Theta$ is calculated directly from $\Omega(s)$ given by Eq.~(\ref{eq5}) and (\ref{eq8}), 
models do not converge well  
if the ratio of the rotational energy to the gravitational energy ($T/W$)
becomes rather high ($\gsim 0.07$). 
This is due to the complexity of the prescribed rotation law. 
Usually, our numerical code
works very efficiently even when $T/W > 0.2$ if the rotation law 
is a well defined function of position as in the case of $v$-constant rotation
or $j$-constant rotation (see Eqs.~\ref{eq17} and~\ref{eq18} below).
We thus pre-determine the specific angular momentum $\hat{j}(s)$ from a
trial density distribution before starting the SCF iteration such that
$\hat{j}(\hat{s}) = \hat{\Omega}(\hat{s})\hat{s}^2$, where $\hat{\Omega}$
is obtained from Eq.~(\ref{eq5}) and (\ref{eq8}).
We then introduce a new rotation law $\hat{\omega}(\hat{s})$ such that 
$\hat{\omega}(\hat{s}) = \hat{j}(\hat{s})/\hat{s}^2$. 
Although $\hat{\Omega}$ and $\hat{\omega}$ are identical in the beginning of the SCF iteration,
the two deviate each other during the SCF iteration, since $\hat{\Omega}$
depends on both density and position (see Eq.~\ref{eq5} and \ref{eq6}) 
while $\hat{\omega}$, given by the pre-determined $\hat{j}(\hat{s})$, 
depends only on position.
Once the SCF integration is completed,
we calculate again  $\hat{\Omega}$ by the use of Eq.~(\ref{eq5}) 
and (\ref{eq8}) from the converged model,
and determine a new $\hat{j}(\hat{s})$. Then, the SCF iteration is 
carried out again with the new $\hat{j}(\hat{s})$. 
This procedure is repeated until 
$\hat{\Omega}$ and $\hat{\omega}$ converge to each other, within an accuracy
of $3 \%$.

\subsection{Model sequences}

\begin{table*}[th]
\begin{center}
\caption{Properties of selected models in sequence AA. 
The columns has the following meaning.
$M$: white dwarf mass, 
$J$: total angular momentum, 
$\Omega_\mathrm{c}$: central angular velocity, 
$\Omega_\mathrm{m}$: moment-of-inertia-weighted mean of angular velocity,
$W$: gravitational energy,  
$T$: rotational energy, 
$U$: internal energy, 
$T/W$ : ratio of the rotational energy to the gravitational energy, 
$R_\mathrm{e}$ : equatorial radius, 
$R_\mathrm{p}/R_\mathrm{e}$: ratio of the polar radius
to the equatorial radius.
}\label{tab1} 
\vspace{0.3cm}
\begin{tabularx}{\linewidth}{ >{\centering\arraybackslash}X  >{\centering\arraybackslash}X >{\centering\arraybackslash}X %
>{\centering\arraybackslash}X  >{\centering\arraybackslash}X  >{\centering\arraybackslash}X %
>{\centering\arraybackslash}X >{\centering\arraybackslash}X  >{\centering\arraybackslash}X  %
>{\centering\arraybackslash}X >{\centering\arraybackslash}X}
\hline \hline
No. &
$M$ & $J$   &$\Omega_\mathrm{c}$& $\Omega_\mathrm{m}$ &  $W$  & $T$ & %
$U$ & $T/W$ & $R_\mathrm{e}$ & $R_\mathrm{p}/R_\mathrm{e}$ \\

& $[\mathrm{M}_\odot]$&$[10^{50}~\mathrm{erg\, s}]$& $[\mathrm{rad/s}]$& $[\mathrm{rad/s}]$  %
&$[10^{50}~\mathrm{erg}]$&
$[10^{50}~\mathrm{erg}]$&[$10^{50}~\mathrm{erg}$]& & 
$[0.01 R_\odot]$&  \\
\hline
\multicolumn{11}{c}{Sequence AA:~ $f_\mathrm{sh} = 1.0$ ;~ $f_\mathrm{K}=0.95$ ;~ $f_\mathrm{p}=0.1$ ;~ $a=1.2$} \\
\hline
\multicolumn{11}{c}{$\rho_\mathrm{c} = 1\times10^8~\mathrm{g\, cm^{-3}}$} \\
\hline
 AAa8 & 1.301 &  0.875 &  0.181 &  0.881 &  9.848 &  0.393 &  6.246 &  0.040 &  0.959 &  0.609 \\
 AAa13 & 1.351 &  1.095 &  0.310 &  0.964 & 10.351 &  0.536 &  6.386 &  0.052 &  1.026 &  0.558 \\
 AAa16 & 1.391 &  1.267 &  0.387 &  1.008 & 10.746 &  0.650 &  6.493 &  0.061 &  1.077 &  0.529 \\
 AAa18 & 1.420 &  1.398 &  0.439 &  1.031 & 11.037 &  0.736 &  6.569 &  0.067 &  1.117 &  0.500 \\
 AAa20 & 1.455 &  1.561 &  0.491 &  1.051 & 11.380 &  0.841 &  6.654 &  0.074 &  1.167 &  0.478 \\
 AAa22 & 1.493 &  1.730 &  0.542 &  1.070 & 11.765 &  0.953 &  6.758 &  0.081 &  1.211 &  0.457 \\
 AAa23 & 1.530 &  1.911 &  0.568 &  1.075 & 12.120 &  1.064 &  6.841 &  0.088 &  1.264 &  0.428 \\
 AAa24 & 1.557 &  2.043 &  0.594 &  1.081 & 12.396 &  1.148 &  6.910 &  0.093 &  1.294 &  0.420 \\
 AAa25 & 1.584 &  2.171 &  0.620 &  1.086 & 12.668 &  1.231 &  6.980 &  0.097 &  1.322 &  0.406 \\
 AAa26 & 1.619 &  2.355 &  0.646 &  1.083 & 13.003 &  1.339 &  7.055 &  0.103 &  1.377 &  0.391 \\
 AAa27 & 1.656 &  2.540 &  0.671 &  1.085 & 13.384 &  1.454 &  7.150 &  0.109 &  1.408 &  0.377 \\
 AAa28 & 1.697 &  2.770 &  0.697 &  1.072 & 13.755 &  1.578 &  7.224 &  0.115 &  1.476 &  0.355 \\
 AAa29 & 1.756 &  3.097 &  0.749 &  1.066 & 14.358 &  1.771 &  7.365 &  0.123 &  1.522 &  0.341 \\
 AAa30 & 1.804 &  3.369 &  0.775 &  1.056 & 14.825 &  1.922 &  7.469 &  0.130 &  1.574 &  0.326 \\
 AAa31 & 1.869 &  3.736 &  0.788 &  1.051 & 15.512 &  2.137 &  7.630 &  0.138 &  1.612 &  0.319 \\
\hline
\multicolumn{11}{c}{$\rho_\mathrm{c} = 2\times10^9~\mathrm{g\, cm^{-3}}$} \\
\hline
 AAe2 & 1.432 &  0.361 &  0.115 &  2.280 & 30.699 &  0.426 & 24.812 &  0.014 &  0.414 &  0.696 \\
 AAe6 & 1.454 &  0.438 &  0.577 &  2.618 & 31.295 &  0.584 & 25.032 &  0.019 &  0.429 &  0.667 \\
 AAe8 & 1.466 &  0.479 &  0.808 &  2.776 & 31.638 &  0.674 & 25.161 &  0.021 &  0.437 &  0.652 \\
 AAe10 & 1.481 &  0.529 &  1.039 &  2.949 & 32.073 &  0.788 & 25.322 &  0.025 &  0.448 &  0.630 \\
 AAe12 & 1.496 &  0.577 &  1.270 &  3.096 & 32.501 &  0.903 & 25.478 &  0.028 &  0.457 &  0.616 \\
 AAe14 & 1.513 &  0.628 &  1.501 &  3.237 & 32.980 &  1.029 & 25.657 &  0.031 &  0.467 &  0.594 \\
 AAe18 & 1.553 &  0.749 &  1.963 &  3.507 & 34.143 &  1.336 & 26.087 &  0.039 &  0.492 &  0.558 \\
 AAe20 & 1.579 &  0.825 &  2.194 &  3.640 & 34.874 &  1.533 & 26.353 &  0.044 &  0.507 &  0.529 \\
 AAe22 & 1.605 &  0.904 &  2.425 &  3.754 & 35.652 &  1.743 & 26.635 &  0.049 &  0.523 &  0.507 \\
 AAe24 & 1.639 &  1.004 &  2.656 &  3.863 & 36.620 &  2.006 & 26.978 &  0.055 &  0.543 &  0.486 \\
 AAe26 & 1.677 &  1.118 &  2.887 &  3.952 & 37.695 &  2.303 & 27.352 &  0.061 &  0.565 &  0.464 \\
 AAe28 & 1.721 &  1.255 &  3.118 &  4.023 & 38.932 &  2.657 & 27.769 &  0.068 &  0.592 &  0.435 \\
 AAe30 & 1.759 &  1.373 &  3.349 &  4.069 & 40.027 &  2.964 & 28.142 &  0.074 &  0.614 &  0.413 \\
 AAe32 & 1.841 &  1.633 &  3.580 &  4.109 & 42.342 &  3.622 & 28.908 &  0.086 &  0.659 &  0.377 \\
 AAe34 & 1.898 &  1.831 &  3.811 &  4.100 & 43.950 &  4.102 & 29.417 &  0.093 &  0.692 &  0.355 \\
 AAe36 & 1.952 &  2.016 &  4.042 &  4.082 & 45.460 &  4.546 & 29.898 &  0.100 &  0.723 &  0.333 \\
 AAe37 & 2.029 &  2.300 &  4.158 &  4.021 & 47.602 &  5.189 & 30.539 &  0.109 &  0.761 &  0.312 \\
\hline
\multicolumn{11}{c}{$\rho_\mathrm{c} = 4\times10^9~\mathrm{g\, cm^{-3}}$} \\
\hline
 AAf3 & 1.448 &  0.313 &  0.327 &  3.047 & 39.226 &  0.492 & 32.819 &  0.013 &  0.344 &  0.696 \\
 AAf5 & 1.457 &  0.346 &  0.653 &  3.283 & 39.560 &  0.579 & 32.949 &  0.015 &  0.350 &  0.681 \\
 AAf7 & 1.469 &  0.382 &  0.980 &  3.528 & 39.961 &  0.684 & 33.104 &  0.017 &  0.357 &  0.667 \\
 AAf9 & 1.480 &  0.419 &  1.307 &  3.753 & 40.387 &  0.795 & 33.272 &  0.020 &  0.363 &  0.645 \\
 AAf11 & 1.494 &  0.458 &  1.633 &  3.971 & 40.865 &  0.918 & 33.460 &  0.022 &  0.371 &  0.630 \\
 AAf13 & 1.508 &  0.501 &  1.960 &  4.181 & 41.391 &  1.057 & 33.663 &  0.026 &  0.378 &  0.616 \\
 AAf15 & 1.524 &  0.544 &  2.287 &  4.376 & 41.957 &  1.203 & 33.886 &  0.029 &  0.386 &  0.594 \\
 AAf17 & 1.543 &  0.596 &  2.613 &  4.583 & 42.648 &  1.385 & 34.156 &  0.032 &  0.396 &  0.572 \\
 AAf19 & 1.563 &  0.652 &  2.940 &  4.763 & 43.369 &  1.579 & 34.427 &  0.036 &  0.407 &  0.558 \\
 AAf21 & 1.585 &  0.710 &  3.267 &  4.934 & 44.171 &  1.791 & 34.736 &  0.041 &  0.418 &  0.536 \\
 AAf23 & 1.607 &  0.767 &  3.593 &  5.073 & 44.956 &  1.997 & 35.038 &  0.044 &  0.428 &  0.514 \\
 AAf25 & 1.637 &  0.846 &  3.920 &  5.232 & 46.047 &  2.289 & 35.449 &  0.050 &  0.443 &  0.493 \\
 AAf27 & 1.669 &  0.932 &  4.247 &  5.369 & 47.207 &  2.607 & 35.881 &  0.055 &  0.459 &  0.471 \\
 AAf29 & 1.706 &  1.032 &  4.574 &  5.485 & 48.542 &  2.972 & 36.372 &  0.061 &  0.477 &  0.449 \\
 AAf31 & 1.750 &  1.155 &  4.900 &  5.568 & 50.128 &  3.413 & 36.934 &  0.068 &  0.500 &  0.420 \\
 AAf33 & 1.792 &  1.271 &  5.227 &  5.629 & 51.646 &  3.834 & 37.480 &  0.074 &  0.519 &  0.399 \\
 AAf35 & 1.880 &  1.527 &  5.554 &  5.672 & 54.784 &  4.736 & 38.566 &  0.086 &  0.562 &  0.362 \\
 AAf37 & 1.989 &  1.851 &  5.880 &  5.709 & 58.928 &  5.902 & 40.027 &  0.100 &  0.597 &  0.333 \\
\hline
\end{tabularx}
\end{center}
\end{table*}
\begin{table*}[th]
\begin{center}
\caption{
Same as in Table~\ref{tab1} but with sequences ABe, ACe, ADe, AEe and AFe.
}\label{tab2} 
\vspace{0.1cm}
\begin{tabularx}{\linewidth}{ >{\centering\arraybackslash}X  >{\centering\arraybackslash}X >{\centering\arraybackslash}X %
>{\centering\arraybackslash}X  >{\centering\arraybackslash}X  >{\centering\arraybackslash}X %
>{\centering\arraybackslash}X >{\centering\arraybackslash}X  >{\centering\arraybackslash}X  %
>{\centering\arraybackslash}X >{\centering\arraybackslash}X}
\hline \hline
No. &
$M$ & $J$   &$\Omega_\mathrm{c}$& $\Omega_\mathrm{m}$ &  $W$  & $T$ & %
$U$ & $T/W$ & $R_\mathrm{e}$ & $R_\mathrm{p}/R_\mathrm{e}$ \\

& $[\mathrm{M}_\odot]$&$[10^{50}~\mathrm{erg\, s}]$& $[\mathrm{rad/s}]$& $[\mathrm{rad/s}]$  %
&$[10^{50}~\mathrm{erg}]$&
$[10^{50}~\mathrm{erg}]$&[$10^{50}~\mathrm{erg}$]& & 
$[0.01 R_\odot]$&  \\
\hline
\multicolumn{11}{c}{Sequence ABe:~ $f_\mathrm{sh} = 1.0$ ;~ $f_\mathrm{K}=0.95$ ;~ $f_\mathrm{p}=0.05$ ;~ $a=1.2$ ;~$\rho_\mathrm{c} = 2\times10^9~\mathrm{g~cm^{-3}}$ } \\
\hline
 ABe2 & 1.467 &  0.483 &  0.635 &  2.766 & 31.615 &  0.682 & 25.124 &  0.022 &  0.442 &  0.645 \\
 ABe10 & 1.502 &  0.597 &  1.097 &  3.105 & 32.556 &  0.942 & 25.448 &  0.029 &  0.471 &  0.594 \\
 ABe19 & 1.554 &  0.763 &  1.617 &  3.437 & 33.955 &  1.337 & 25.920 &  0.039 &  0.515 &  0.536 \\
 ABe23 & 1.580 &  0.842 &  1.848 &  3.552 & 34.656 &  1.531 & 26.159 &  0.044 &  0.534 &  0.507 \\
 ABe33 & 1.685 &  1.175 &  2.425 &  3.780 & 37.397 &  2.325 & 27.040 &  0.062 &  0.617 &  0.428 \\
 ABe37 & 1.757 &  1.412 &  2.656 &  3.821 & 39.262 &  2.876 & 27.622 &  0.073 &  0.670 &  0.384 \\
 ABe42 & 1.857 &  1.775 &  2.945 &  3.719 & 41.633 &  3.624 & 28.280 &  0.087 &  0.758 &  0.333 \\
 ABe48 & 2.034 &  2.486 &  3.407 &  3.438 & 45.757 &  4.945 & 29.375 &  0.108 &  0.893 &  0.275 \\
\hline
\multicolumn{11}{c}{Sequence ACe:~ $f_\mathrm{sh} = 1.0$ ;~ $f_\mathrm{K}=0.95$ ;~ $f_\mathrm{p}=0.1$ ;~ $a=1.0$ ;~$\rho_\mathrm{c} = 2\times10^9~\mathrm{g~cm^{-3}}$ } \\
\hline
 ACe5 & 1.457 &  0.450 &  0.462 &  2.631 & 31.321 &  0.607 & 25.011 &  0.019 &  0.451 &  0.638 \\
 ACe13 & 1.516 &  0.642 &  1.386 &  3.226 & 32.973 &  1.048 & 25.612 &  0.032 &  0.492 &  0.565 \\
 ACe19 & 1.579 &  0.833 &  2.079 &  3.596 & 34.765 &  1.528 & 26.267 &  0.044 &  0.533 &  0.507 \\
 ACe25 & 1.678 &  1.133 &  2.772 &  3.894 & 37.553 &  2.298 & 27.243 &  0.061 &  0.597 &  0.435 \\
 ACe29 & 1.753 &  1.366 &  3.234 &  3.997 & 39.697 &  2.895 & 27.981 &  0.073 &  0.641 &  0.399 \\
 ACe32 & 1.849 &  1.681 &  3.580 &  4.028 & 42.375 &  3.671 & 28.858 &  0.087 &  0.700 &  0.355 \\
 ACe35 & 1.969 &  2.102 &  3.927 &  3.973 & 45.693 &  4.654 & 29.893 &  0.102 &  0.771 &  0.312 \\
\hline
\multicolumn{11}{c}{Sequence ADe:~ $f_\mathrm{sh} = 0.1$ ;~ $f_\mathrm{K}=0.95$ ;~ $f_\mathrm{p}=0.20$ ;~ $a=1.2$~;%
~$\rho_\mathrm{c} = 2\times10^9~\mathrm{g\, cm^{-3}}$} \\
\hline
 ADe2 & 1.447 &  0.418 &  0.577 &  2.539 & 31.137 &  0.538 & 24.982 &  0.017 &  0.427 &  0.674 \\
 ADe6 & 1.597 &  0.868 &  2.887 &  3.783 & 35.646 &  1.685 & 26.746 &  0.047 &  0.502 &  0.529 \\
 ADe7 & 1.664 &  1.060 &  3.465 &  4.041 & 37.754 &  2.230 & 27.560 &  0.059 &  0.534 &  0.486 \\
 ADe8 & 1.740 &  1.276 &  4.042 &  4.221 & 40.100 &  2.849 & 28.438 &  0.071 &  0.565 &  0.442 \\
 ADe9 & 1.869 &  1.667 &  4.620 &  4.389 & 44.189 &  3.964 & 29.923 &  0.090 &  0.622 &  0.384 \\
 ADe10 & 2.044 &  2.233 &  5.197 &  4.464 & 49.892 &  5.547 & 31.919 &  0.111 &  0.688 &  0.333 \\
\hline
\multicolumn{11}{c}{Sequence AEe:~ $f_\mathrm{sh} = 0.5$ ;~ $f_\mathrm{K}=0.5$ ;~ $f_\mathrm{p}=0.10$ ;~ $a=1.2$~;%
~$\rho_\mathrm{c} = 2\times10^9~\mathrm{g\, cm^{-3}}$} \\
\hline
 AEe1 & 1.391 &  0.170 &  0.000 &  1.224 & 29.624 &  0.114 & 24.467 &  0.004 &  0.320 &  0.920 \\
 AEe2 & 1.422 &  0.319 &  1.155 &  2.132 & 30.530 &  0.342 & 24.822 &  0.011 &  0.331 &  0.870 \\
 AEe3 & 1.480 &  0.515 &  2.310 &  3.000 & 32.292 &  0.784 & 25.523 &  0.024 &  0.350 &  0.797 \\
 AEe4 & 1.578 &  0.792 &  3.465 &  3.744 & 35.252 &  1.538 & 26.673 &  0.044 &  0.377 &  0.703 \\
 AEe5 & 1.760 &  1.308 &  4.620 &  4.331 & 40.818 &  3.028 & 28.716 &  0.074 &  0.423 &  0.587 \\
 AEe6 & 1.941 &  1.852 &  5.197 &  4.531 & 46.482 &  4.598 & 30.694 &  0.099 &  0.460 &  0.514 \\
\hline
\multicolumn{11}{c}{Sequence AFe:~ $f_\mathrm{sh} = 0.1$ ;~ $f_\mathrm{K}=0.95$ ;~ $f_\mathrm{p}=0.10$ ;~ $a=1.2$~;%
~$\rho_\mathrm{c} = 2\times10^9~\mathrm{g\, cm^{-3}}$} \\
\hline
 AFe2 & 1.393 &  0.175 &  0.577 &  1.242 & 29.657 &  0.136 & 24.459 &  0.005 &  0.379 &  0.775 \\
 AFe6 & 1.481 &  0.526 &  2.887 &  2.990 & 32.231 &  0.787 & 25.469 &  0.024 &  0.438 &  0.638 \\
 AFe8 & 1.582 &  0.825 &  4.042 &  3.723 & 35.252 &  1.567 & 26.630 &  0.044 &  0.495 &  0.536 \\
 AFe9 & 1.664 &  1.066 &  4.620 &  4.017 & 37.684 &  2.222 & 27.519 &  0.059 &  0.542 &  0.478 \\
 AFe10 & 1.784 &  1.431 &  5.197 &  4.203 & 41.208 &  3.201 & 28.748 &  0.078 &  0.610 &  0.406 \\
 AFe11 & 1.939 &  1.941 &  5.775 &  4.226 & 45.746 &  4.501 & 30.247 &  0.098 &  0.689 &  0.348 \\
 AFe12 & 2.046 &  2.318 &  6.006 &  4.190 & 48.912 &  5.439 & 31.263 &  0.111 &  0.741 &  0.312 \\
\hline
\multicolumn{11}{c}{Sequence AGe:~ $f_\mathrm{sh} = -1.0$ ;~ $f_\mathrm{K}=0.6$ ;~ $f_\mathrm{p}=0.01$ ;~ $a=0.9$~;%
~$\rho_\mathrm{c} = 2\times10^9~\mathrm{g\, cm^{-3}}$} \\
\hline
 AGe1 & 1.524 &  0.630 &  5.775 &  3.448 & 33.996 &  1.121 & 26.391 &  0.033 &  0.365 &  0.732 \\
 AGe3 & 1.585 &  0.811 &  6.237 &  3.838 & 35.821 &  1.606 & 27.069 &  0.045 &  0.396 &  0.659 \\
 AGe5 & 1.657 &  1.032 &  6.699 &  4.073 & 37.862 &  2.181 & 27.764 &  0.058 &  0.438 &  0.580 \\
 AGe6 & 1.730 &  1.269 &  6.930 &  4.127 & 39.786 &  2.754 & 28.362 &  0.069 &  0.486 &  0.514 \\
 AGe7 & 1.803 &  1.527 &  7.045 &  4.036 & 41.535 &  3.306 & 28.845 &  0.080 &  0.541 &  0.457 \\
 AGe8 & 1.890 &  1.881 &  7.161 &  3.771 & 43.371 &  3.934 & 29.262 &  0.091 &  0.616 &  0.391 \\
\hline
\multicolumn{11}{c}{Sequence Re:~ Rigid Rotation; ~$\rho_\mathrm{c} = 2\times10^9~\mathrm{g\, cm^{-3}}$} \\
\hline
 Re1 & 1.376 &  0.000 &  0.000 &  0.000 & 29.248 &  0.000 & 24.356 &  0.000 &  0.296 &  1.000 \\
 Re4 & 1.392 &  0.180 &  1.309 &  1.309 & 29.715 &  0.118 & 24.540 &  0.004 &  0.306 &  0.957 \\
 Re6 & 1.415 &  0.294 &  2.004 &  2.004 & 30.399 &  0.294 & 24.802 &  0.010 &  0.326 &  0.884 \\
 Re7 & 1.428 &  0.344 &  2.269 &  2.269 & 30.765 &  0.391 & 24.938 &  0.013 &  0.340 &  0.848 \\
 Re8 & 1.440 &  0.388 &  2.472 &  2.472 & 31.095 &  0.480 & 25.059 &  0.015 &  0.357 &  0.797 \\
 Re9 & 1.449 &  0.422 &  2.613 &  2.613 & 31.355 &  0.551 & 25.151 &  0.018 &  0.377 &  0.754 \\
 Re10 & 1.455 &  0.443 &  2.694 &  2.694 & 31.519 &  0.597 & 25.208 &  0.019 &  0.400 &  0.710 \\
 Re11 & 1.458 &  0.451 &  2.724 &  2.724 & 31.579 &  0.615 & 25.228 &  0.019 &  0.426 &  0.667 \\
\hline
\end{tabularx}
\end{center}
\end{table*}

Hereafter, 
we will refer to the rotation law given by Eq.~(\ref{eq5}) and Eq~(\ref{eq8})  to 
\emph{accreting-white-dwarf rotation} (AWD rotation). 
For comparison, we employ rigid, $v$-constant and $j$-constant rotation laws,
which were also considered by Hachisu (\cite{Hachisu86}).

Model sequences which are constructed in the present study with different parameters
and different rotation laws 
are summarized as follows (see also Tables~1 and~2, below): 
\begin{enumerate}
\item AWD rotation: $\Omega$ given by Eq.~(\ref{eq5}) and~(\ref{eq8})
  \begin{itemize}
     \item Sequence AA: $f_\mathrm{sh}=1.0$, $f_\mathrm{K} = 0.95$, $f_\mathrm{p}=0.10$, $a=1.2$
     \item Sequence AB: $f_\mathrm{sh}=1.0$, $f_\mathrm{K} = 0.95$, $f_\mathrm{p}=0.05$, $a=1.2$
     \item Sequence AC: $f_\mathrm{sh}=1.0$, $f_\mathrm{K} = 0.95$, $f_\mathrm{p}=0.10$, $a=1.0$
     \item Sequence AD: $f_\mathrm{sh}=1.0$, $f_\mathrm{K} = 0.95$, $f_\mathrm{p}=0.20$, $a=1.2$
     \item Sequence AE: $f_\mathrm{sh}=0.5$, $f_\mathrm{K} = 0.50$, $f_\mathrm{p}=0.10$, $a=1.2$
     \item Sequence AF: $f_\mathrm{sh}=0.1$, $f_\mathrm{K} = 0.95$, $f_\mathrm{p}=0.10$, $a=1.2$
     \item Sequence AG: $f_\mathrm{sh}=-1.0$, $f_\mathrm{K} = 0.6$, $f_\mathrm{p}=0.01$, $a=0.9$
  \end{itemize}
\item rigid rotation (Sequence R): $\Omega =$ constant
\item $j$-constant rotation (Sequence J): 
\begin{equation}\label{eq17}
\Omega = j_\mathrm{0}/(0.01R_\mathrm{e}^2+s^2)~.
\end{equation}
\item $v$-constant rotation (Sequence V): 
\begin{equation}\label{eq18}
\Omega = v_\mathrm{0}/(0.01R_\mathrm{e}^2 + s^2)^{1/2}~.
\end{equation}
\end{enumerate}

Here, sequences AA, AB \& AC  are intended for describing the properties 
of 1-D models in Paper~I.
Sequences AD, AE \& AF are to consider
uncertainties due to changes of the given parameters (see below).
Sequence AG is to investigate 
the white dwarf structure 
when the inner core rotates at the shear rate of $-\sigma_\mathrm{DSI, crit}$.

We consider model sequences at 7 different values of $\rho_\mathrm{max}$: 1.0, 2.0, 5.0, 8.0, 20, 
40.0, and 100 $\times 10^{8}~\mathrm{g~cm^{-3}}$. 
Every sequence is given a subindex (a, b, c, d, e, f and g), in order to 
indicate the corresponding $\rho_\mathrm{max}$.
For instance, AAa denotes the sequence with
$f_\mathrm{sh}=1.0$, $f_\mathrm{K} = 0.95$, $f_\mathrm{p}=0.10$, $a=1.2$, and with 
$\rho_\mathrm{max}=1.0\times10^{8}~\mathrm{g~cm^{-3}}$, etc. 
Note that in all models with the AWD- or rigid rotation law 
the maximum density $\rho_\mathrm{max}$
corresponds to the central density $\rho_\mathrm{c}$. 
However, they may differ from each other for $v$-constant
and $j$-constant rotation laws when the ratio
of the rotational energy to the gravitational energy ($T/W$) becomes sufficiently high.
\begin{figure*}
\center
\resizebox{0.45\hsize}{!}{\includegraphics{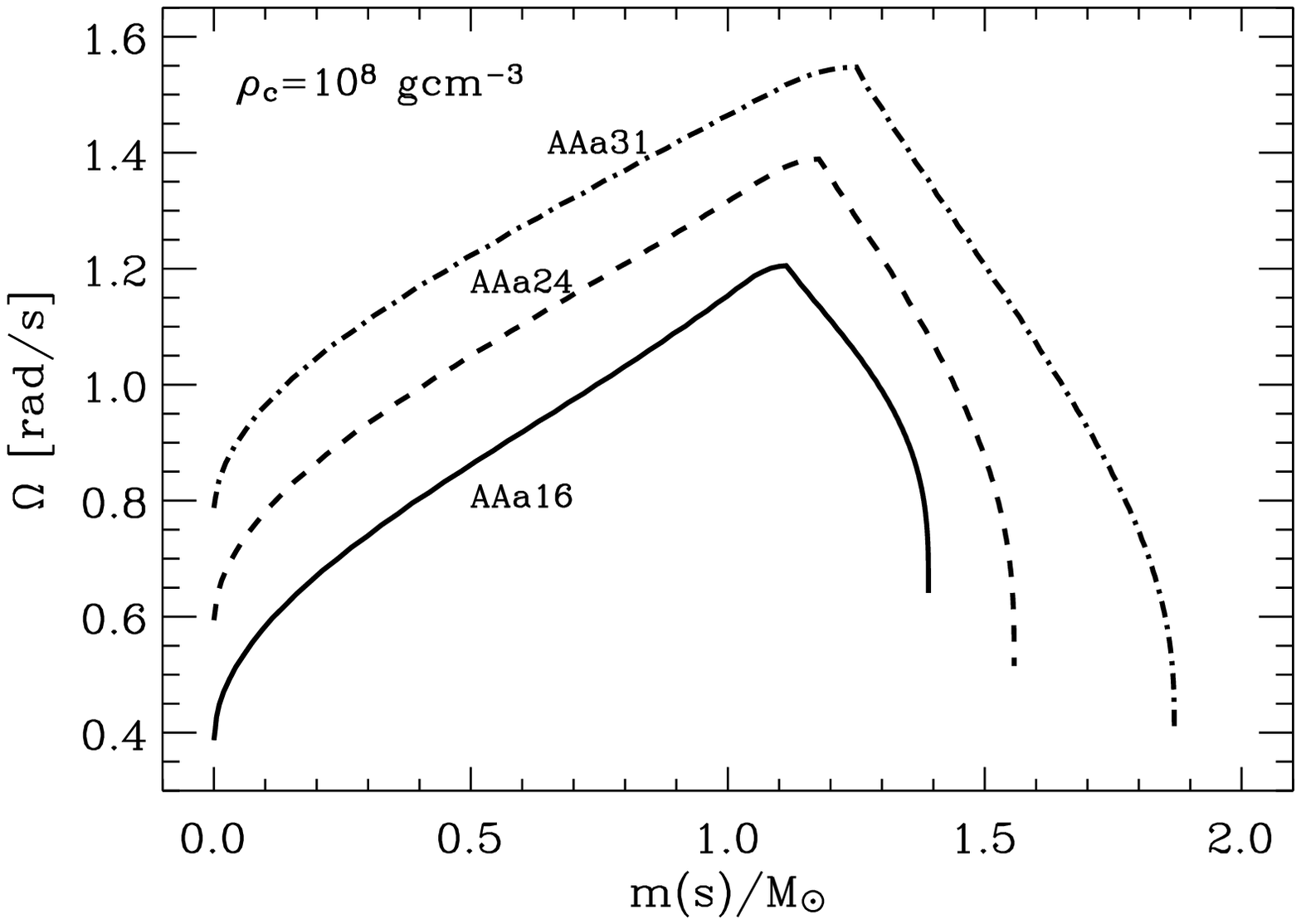}}
\resizebox{0.45\hsize}{!}{\includegraphics{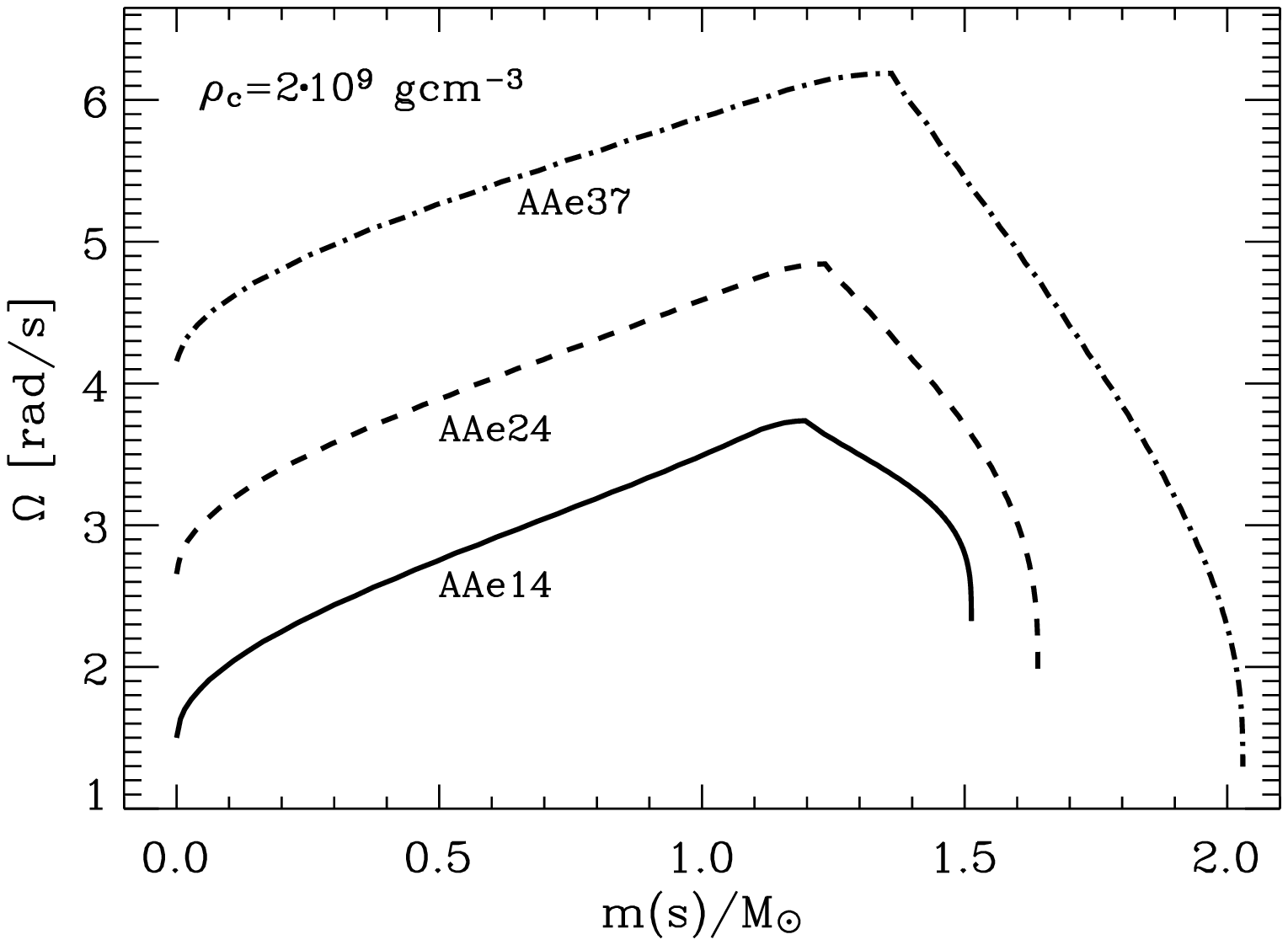}}
\resizebox{0.45\hsize}{!}{\includegraphics{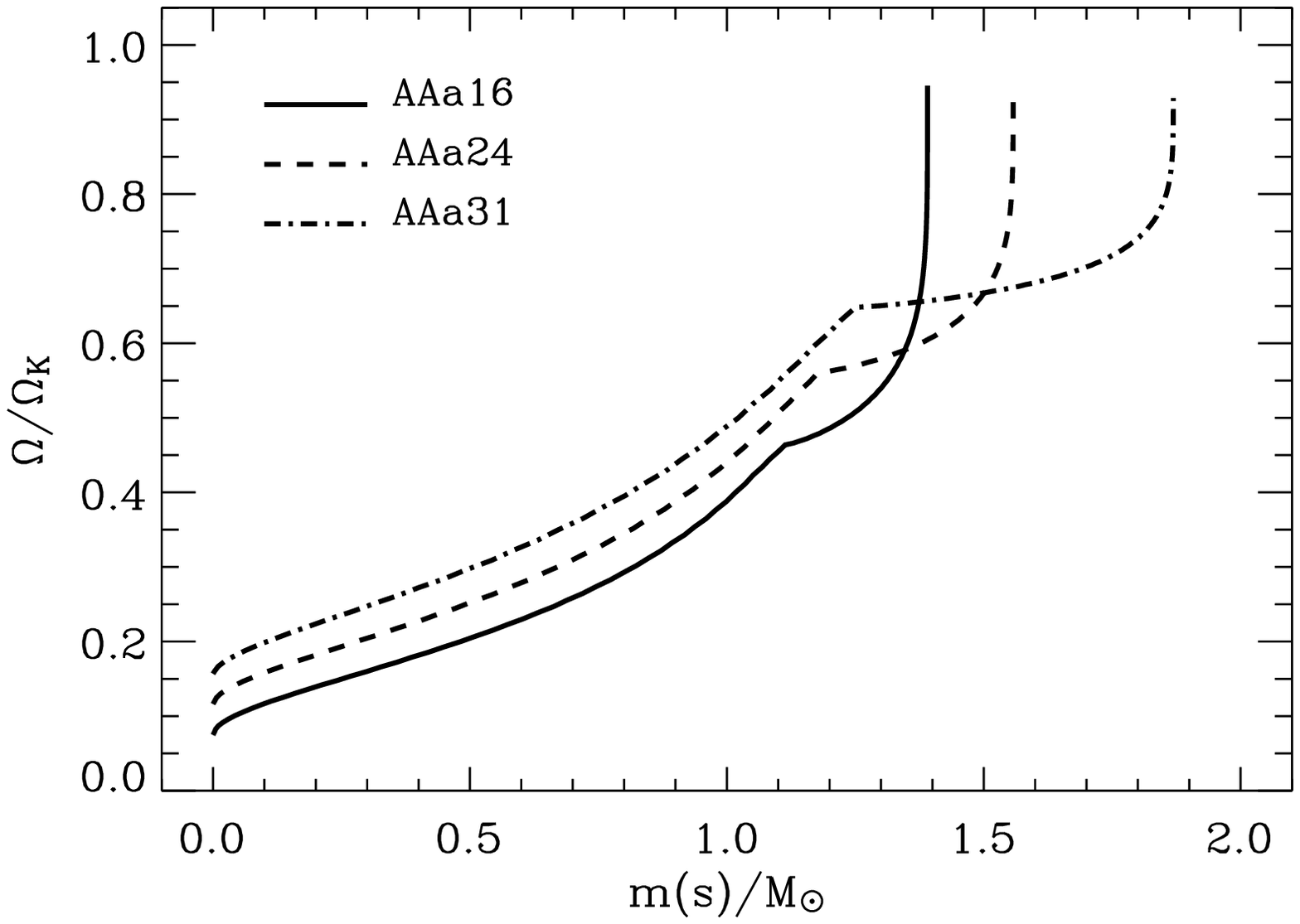}}
\resizebox{0.45\hsize}{!}{\includegraphics{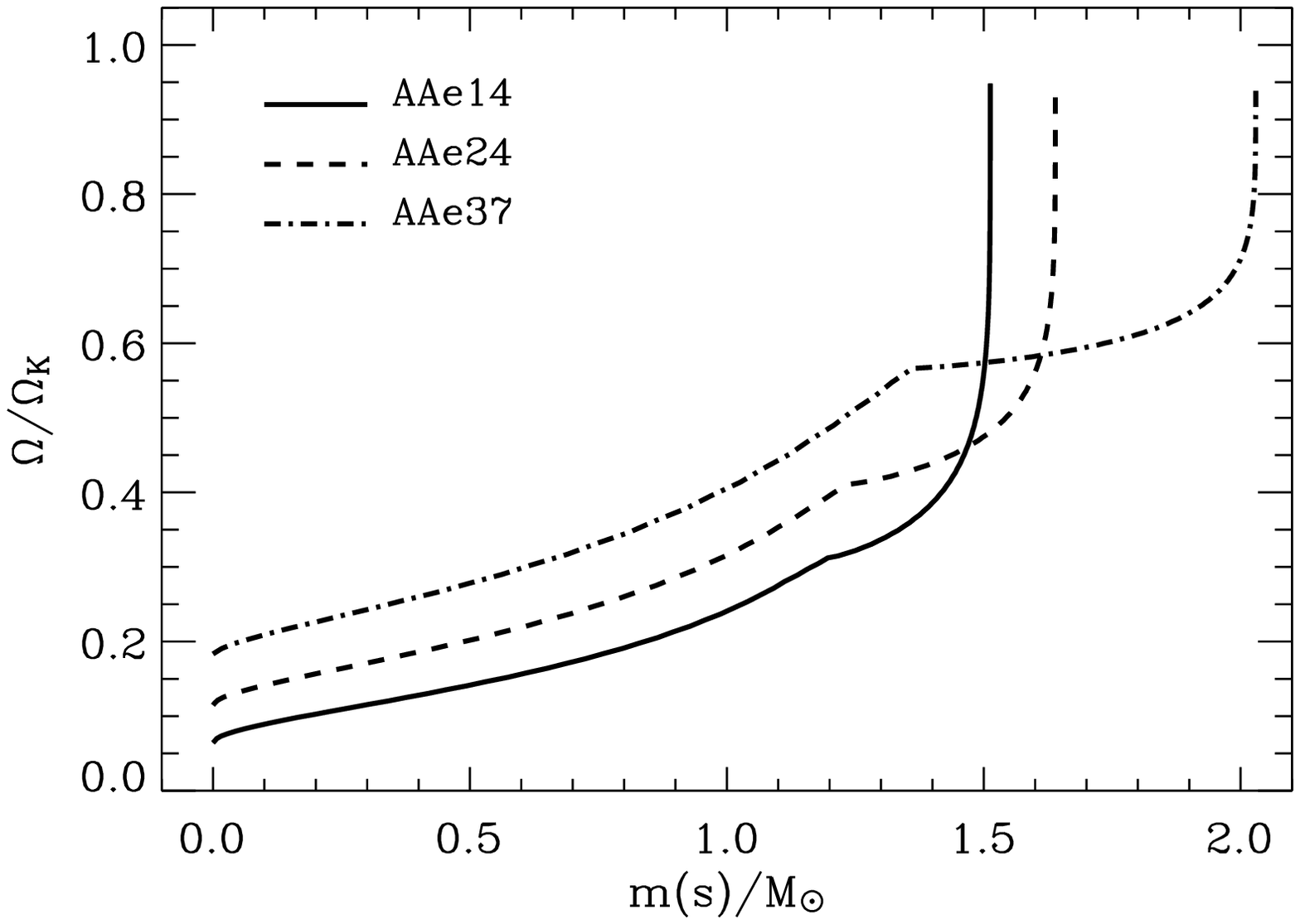}}
\caption{
\emph{Upper panels}: Angular velocity as a function of the mass coordinate
in white dwarf models (AAa16, AAa24 \& AAa31 in the left panel, 
and  AAe10, AAe24 \& AAe37 in the right panel).
~\emph{Lower panels}: Angular velocity normalized to the local Keplerian value
as a function of the mass coordinate, in the corresponding models.
}\label{fig:omega}
\end{figure*}
Within each sequence, white dwarf models 
with different values of $\Omega_\mathrm{c}$ are constructed.  
Since the mass budget in binary systems which are considered
for the Single Degenerate scenario for SNe~Ia is limited, 
the maximum possible mass that a white dwarf can achieve by mass accretion
could not significantly exceed $2.0~\mathrm{M}_\odot$ (Langer et al.~\cite{Langer00}).
Therefore, we do not construct models with $M \gsim 2.1~\mathrm{M}_\odot$.

\section{Basic model properties}\label{sect:results}

Properties of selected models with the AWD rotation law
are presented in Tables~\ref{tab1} and~\ref{tab2}.
As mentioned earlier,  
$\Omega$ profiles in the models with $f_\mathrm{sh} = 1.0$, 
$f_\mathrm{p} = 0.05~\mathrm{or}~0.1$, 
and $f_\mathrm{K} = 0.95$ (Seqs. AA, AB, \& AC)
resemble closest the 1-D models in Paper~I.
The models with $\rho_\mathrm{c}=2\times10^9~\mathrm{g\, cm^{-3}}$
represent the pre-explosion stages, since carbon burning is expected to occur
at roughly this density
in accreting white dwarfs with $\dot{M} \gsim 10^{-7}~
\mathrm{M}_\odot~\mathrm{yr}^{-1}$ (e.g. Nomoto~\cite{Nomoto82}). 

Fig.~\ref{fig:omega} shows angular velocity profiles 
of six different models, as indicated.
Here, the mass coordinate $m(s)$ is defined such that
$m(s) = 2\pi\int_{0}^{s}\int_{-\infty}^{\infty} \rho(s',z) dz ds'$.
These profiles have similar shapes to those
in 1-D white models in Paper~I. A detailed comparison
to 1-D models is performed in Sect.~\ref{sect:comparison}.
Note that models with $\rho_\mathrm{c} = 2\times10^9~\mathrm{g~cm^{-3}}$
have higher values of $\Omega$ 
than those with $\rho_\mathrm{c} = 1\times10^8~\mathrm{g~cm^{-3}}$
at a given mass, due to their stronger compactness.
This tendency is reflected 
in the values of the moment-of-inertia weighted mean
of $\Omega$ in Table~\ref{tab1}.

\begin{figure}
\center
\resizebox{\hsize}{!}{\includegraphics{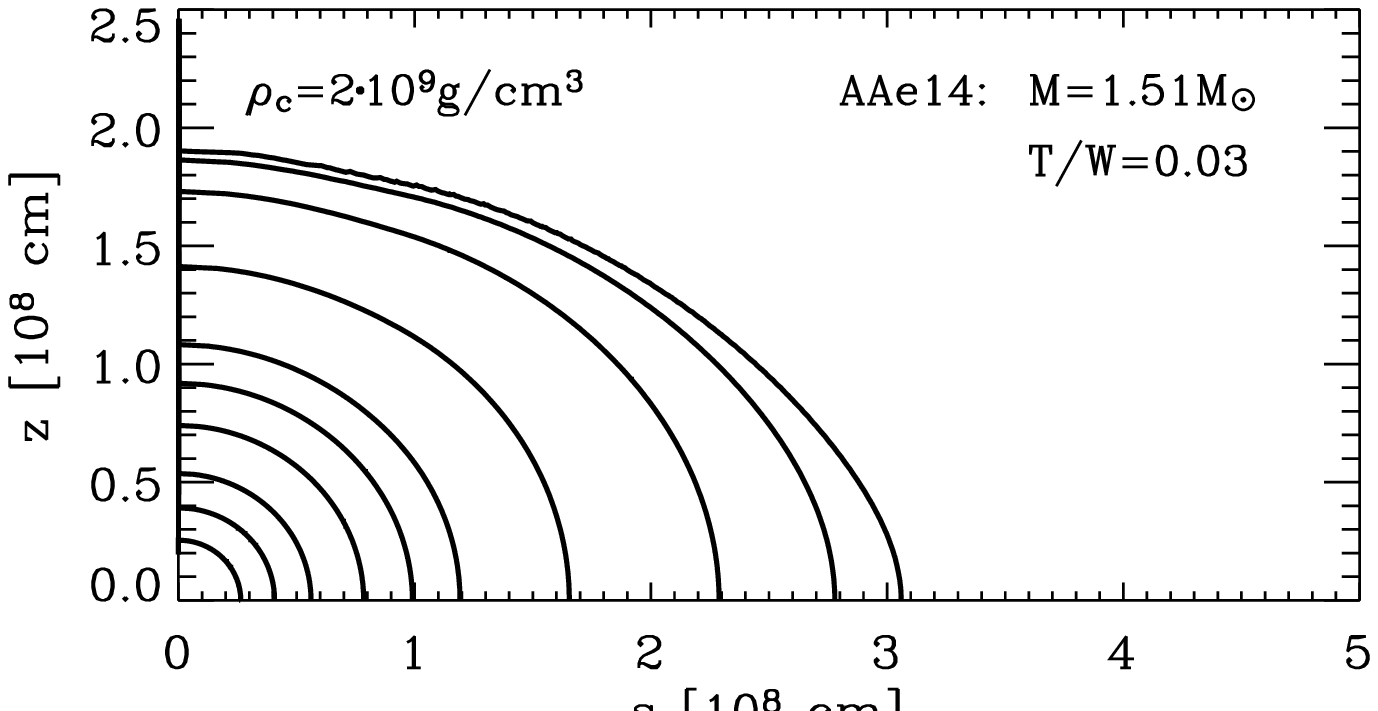}}
\resizebox{\hsize}{!}{\includegraphics{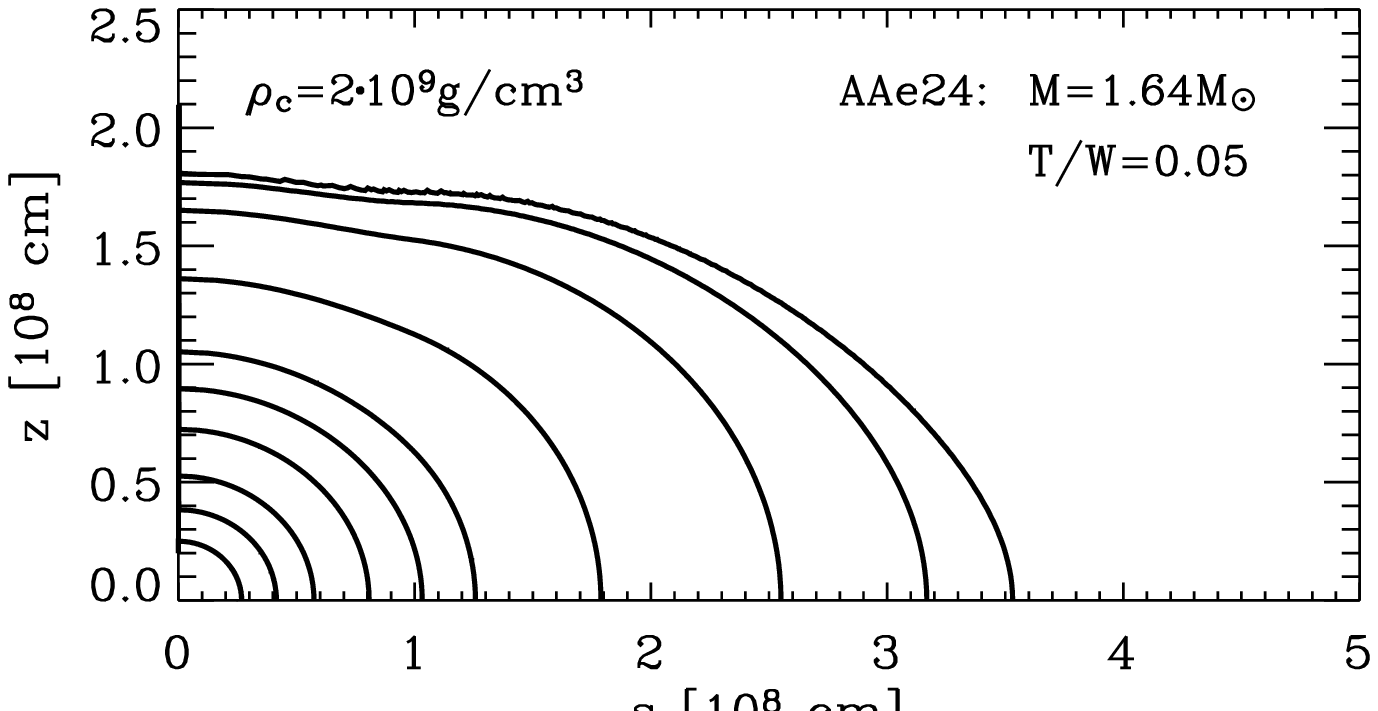}}
\resizebox{\hsize}{!}{\includegraphics{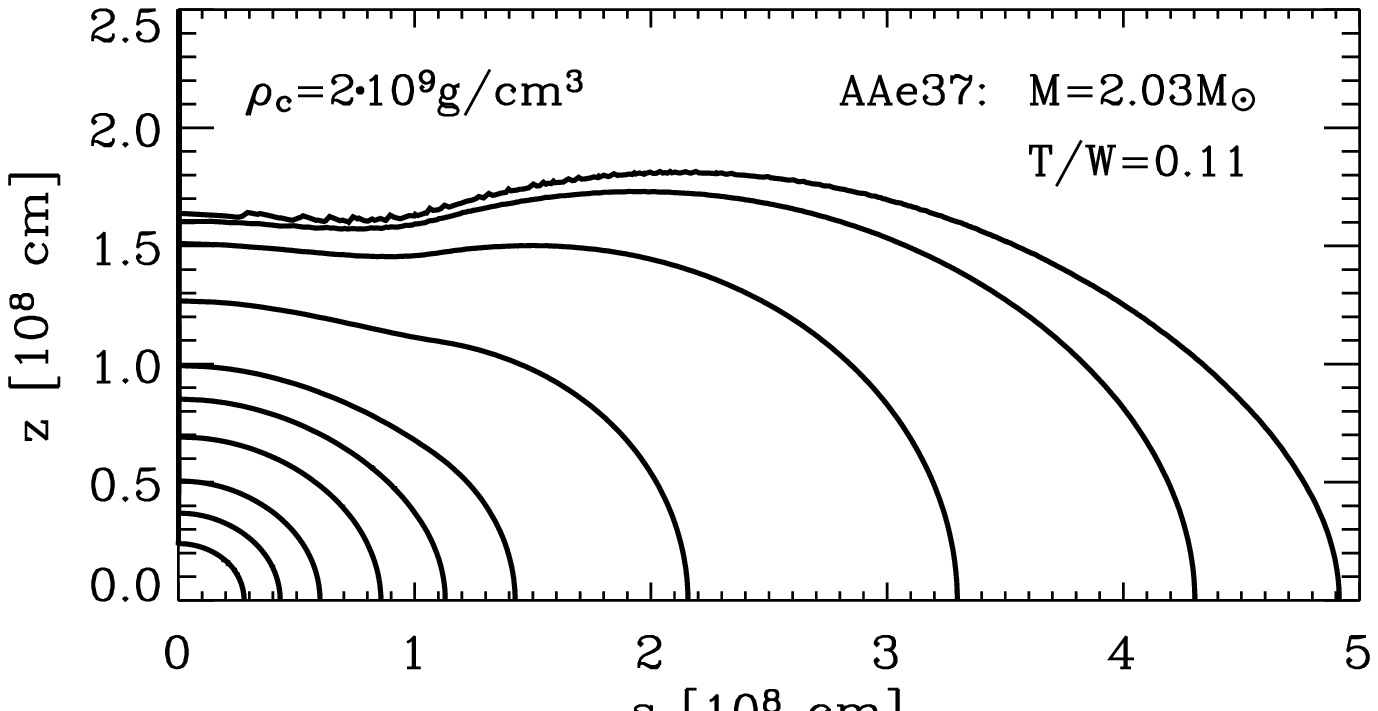}}
\caption{Iso-density contours of white dwarf models AAe14, AAe24 and AAe37. 
These models have the same central density  of
$\rho_\mathrm{c} = 2\times10^9~\mathrm{g~cm^{-3}}$, 
but different masses and angular momenta.
The contour lines denote levels of $0.8\rho_\mathrm{c}$, 
$0.6\rho_\mathrm{c}$, $0.4\rho_\mathrm{c}$, $0.2\rho_\mathrm{c}$, $0.1\rho_\mathrm{c}$, $0.05\rho_\mathrm{c}$, 
$10^{-2}\rho_\mathrm{c}$, $10^{-3}\rho_\mathrm{c}$, $10^{-4}\rho_\mathrm{c}$ and $10^{-5}\rho_\mathrm{c}$,
from the central region to the outer layers.
}\label{fig:contour}
\end{figure}

Iso-density contours of the corresponding models with 
$\rho_\mathrm{c}=2\times10^9~\mathrm{g~cm^{-3}}$ are given in Fig.~\ref{fig:contour}. 
As the value of $T/W$ increases, the white dwarf structure
deviates from the spheroidal shape more and more, 
and the ratio of the polar radius
to the equatorial radius decreases.
However, the inner core with $\rho \gsim 0.05 \rho_\mathrm{c}$, 
which contains more than 75\%  of the total mass,
remains fairly spheroidal even when $T/W = 0.11$. 

\begin{figure}
\center
\resizebox{\hsize}{!}{\includegraphics{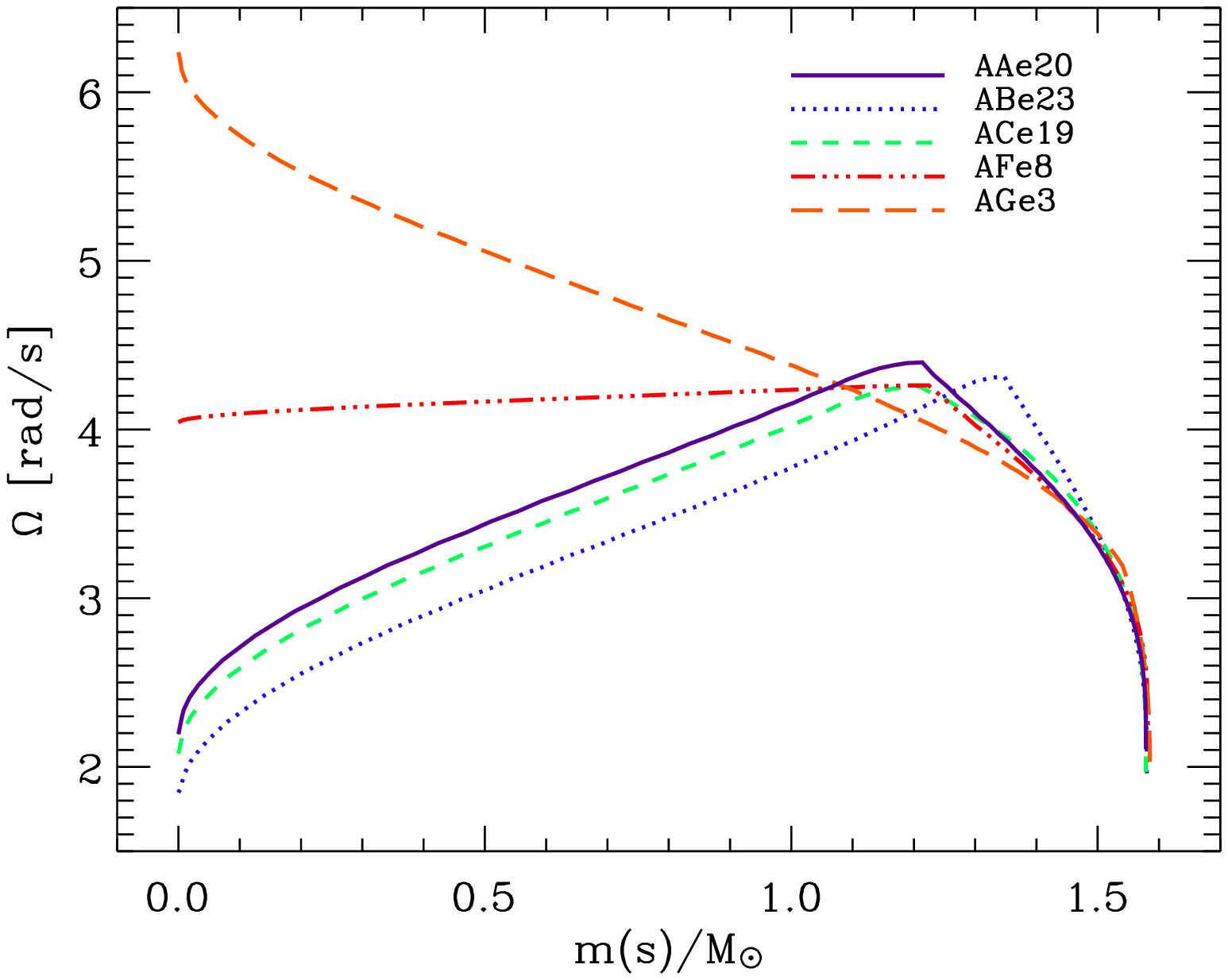}}
\resizebox{\hsize}{!}{\includegraphics{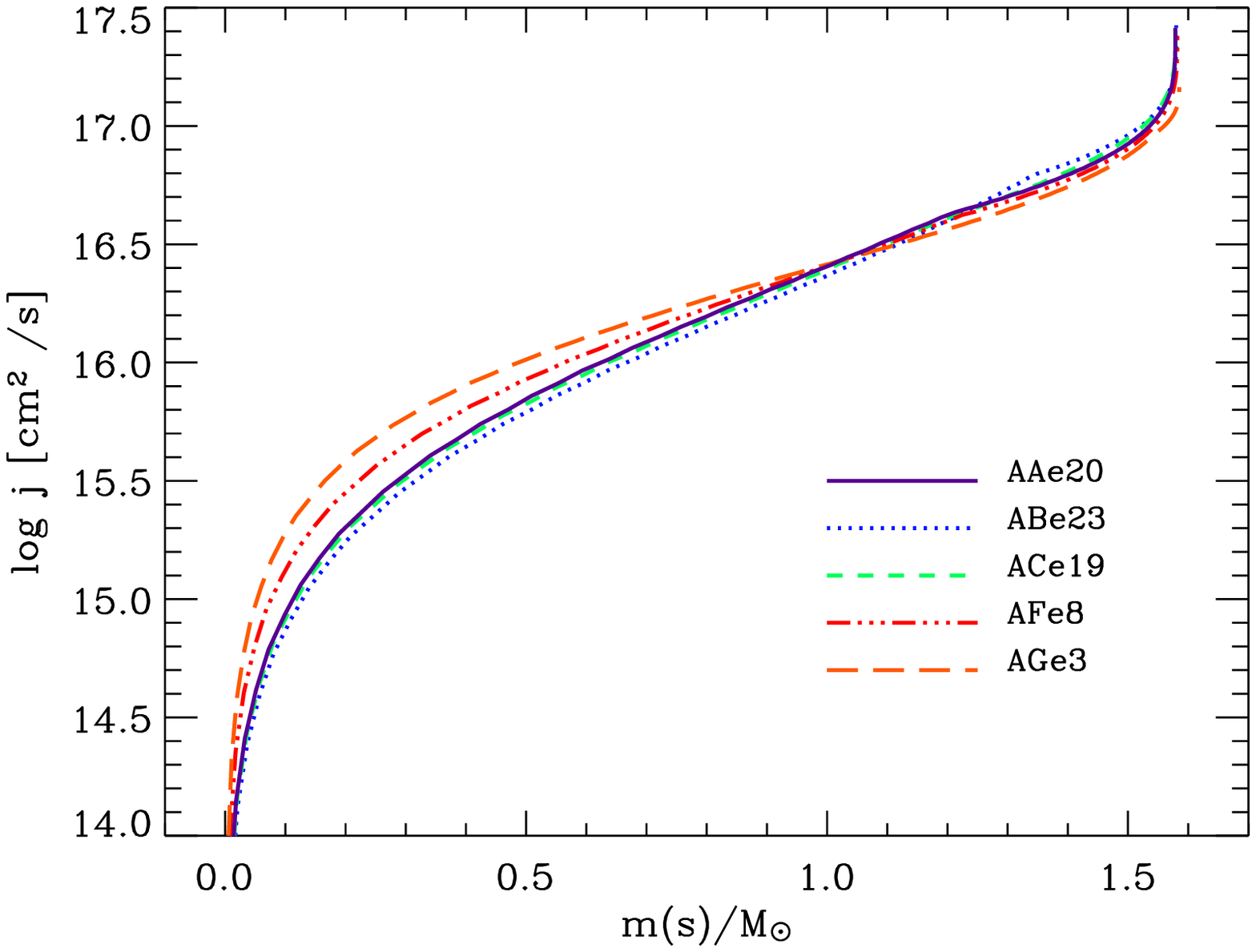}}
\caption{\emph{Upper panel:} Angular velocity as a function of the mass coordinate
in white dwarf models AAe20 (solid line), ABe20 (dotted line), 
ACe19 (short-dashed), AFe8 (dashed-three-dotted line)
and AGe3 (long-dashed line).
\emph{Lower panel:} Specific angular momentum in the corresponding models of the upper panel.
}\label{fig:omega_comp}
\end{figure}

\begin{figure}
\center
\resizebox{\hsize}{!}{\includegraphics{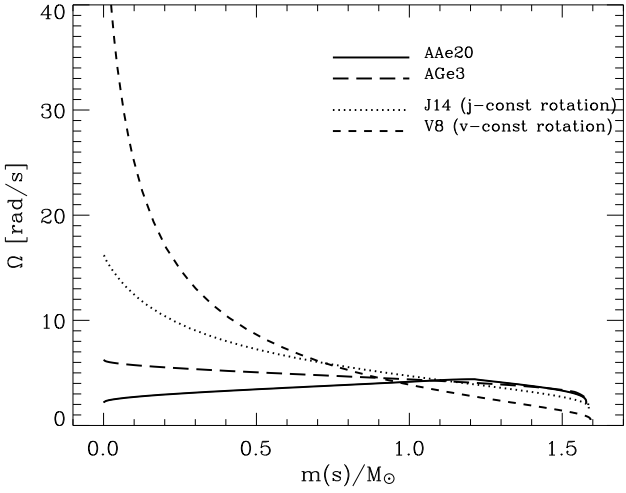}}
\caption{
Angular velocity as a function of the mass-coordinate in models AAe20 (solid line), AGe3 (long-dashed line), 
J14 (dotted line) and V8 (short-dashed line). 
}\label{fig:omega_comp2}
\end{figure}

\begin{figure}
\center
\resizebox{\hsize}{!}{\includegraphics{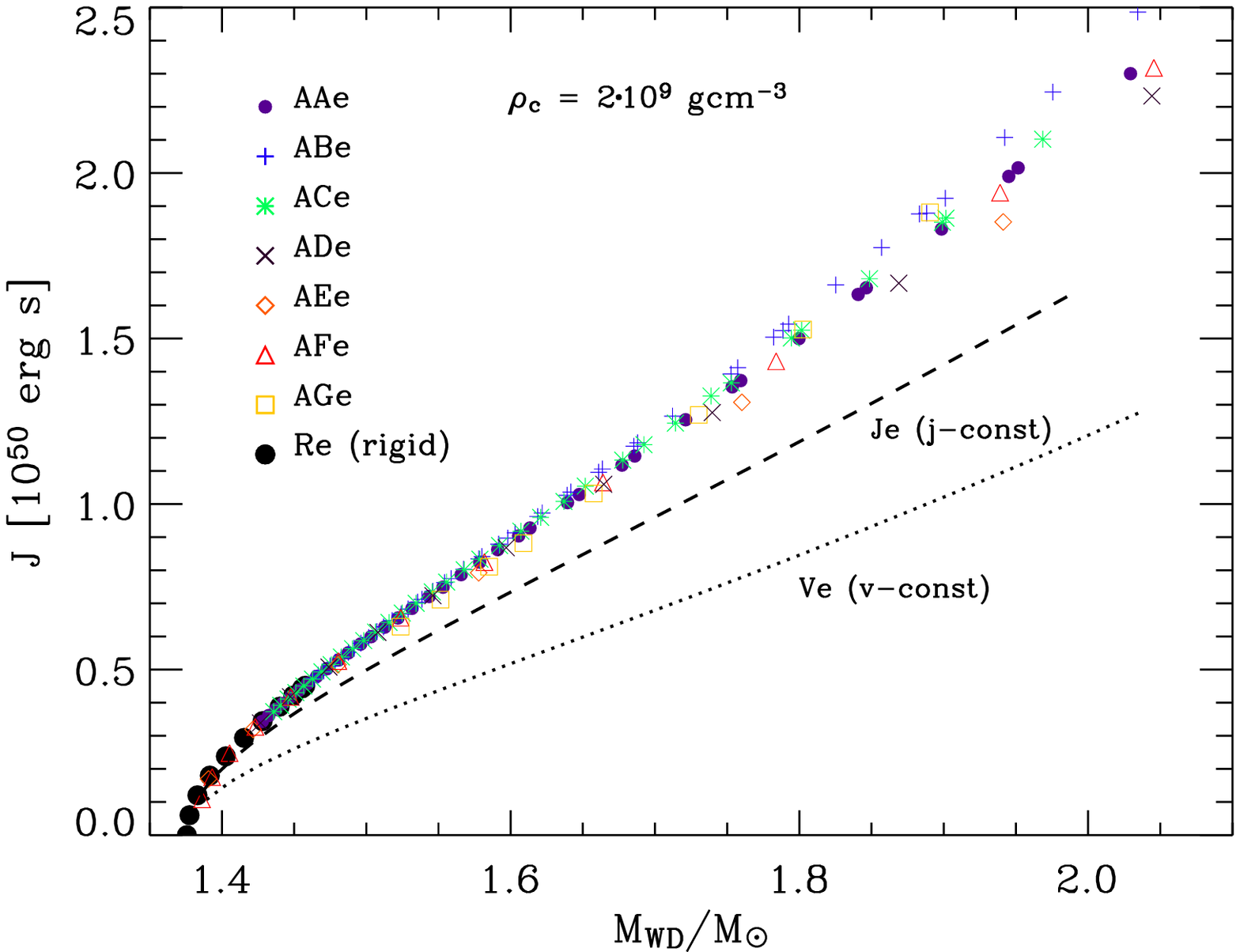}}
\caption{
Total angular momentum of white dwarf models as a function of the 
total mass of the white dwarf, in sequences AAe, ABe, ACe, AFe and AGe.
}\label{fig:mj1}
\end{figure}

Even though our 2-D models are designed to resemble the results from
evolutionary calculations, the parameter range which describes them
is finite due to two effects. Firstly, variations in the evolution itself,
i.e. of the accretion rate in binary systems, or of the initial white dwarf mass,
give rise to variations, say, in the pre-supernova modes (i.e., models
with $\rho_\mathrm{c} = 2\times10^9~\mathrm{g, cm^{-3}}$); i.e., the
value of $f_\mathrm{p}$ has been shown to depend on both in Paper~I. Secondly,
some physical processes described in the evolutionary models are uncertain.
For instance, the exact value of the critical Richardson number is still uncertain, 
and instead of $R_\mathrm{i,c} = 0.25$ that is used in the present study, 
$R_\mathrm{i,c} = 1.0$ is also often adopted 
in the literature (e.g. Hirschi et al.~\cite{Hirschi04}), 
with which the shear rate in the inner core of white dwarfs would
become weaker than assumed in sequence AA. 
Therefore, we constructed 2-D models in which we varied all relevant parameters (Table~\ref{tab2}).

Fig.~\ref{fig:omega_comp} shows $\Omega$-profiles in 5 different models 
(AAe20, ABe23, ACe19, AFe8 \& AG3), 
which have the same central density ($\rho_\mathrm{c} = 2\times10^9~\mathrm{g, cm^{-3}}$) 
and similar masses ($\sim 1.58~\mathrm{M}_\odot$), but different sets of $f_\mathrm{p}$, 
$f_\mathrm{sh}$, $f_\mathrm{K}$ and $a$ (see Tables~\ref{tab1} and~\ref{tab2}). 
A comparison of model AAe20 with model ABe23 shows the influence of $f_\mathrm{p}$. 
For smaller $f_\mathrm{p}$, the $\Omega$-peak is located further outward.
As a result, the inner core rotates more slowly and the outer region
more rapidly, for fixed mass and central density.
The influence of $a$ can be understood by comparing model AAe20 with model ACe19:
the surface region rotates more rapidly and the inner core more slowly in model ACe19.
This is because, with a smaller $a$, 
the role of the centrifugal force becomes stronger in the region of $s > s_\mathrm{p}$
(see Eq.~\ref{eq8}). 
Model AFe8 has a weaker shear rate in the inner core than model AAe20. 
As a consequence, this model needs a higher value of $\Omega_\mathrm{c}$ than model AAe20,
to retain the same central density at a given mass.
In model AGe3, where the $\Omega$-gradient is negative throughout the white dwarf, 
the inner core rotates much faster than the other models. 
In Fig.~\ref{fig:omega_comp2}, models with $j$-constant and $v$-constant rotation laws
are compared with models AAe20 and AGe3. Note that differential rotation in those models is
much stronger than in model AGe30, which means that the shear rates
for such exotic rotation laws are well above the critical value for the onset of 
the dynamical shear instability, $\sigma_\mathrm{DSI,crit}$.  

The specific angular momentum profiles in Fig.~\ref{fig:omega_comp} show an interesting feature. 
If a white dwarf model has lower values of $j$ in the inner core than the other models do, 
it has higher values of $j$ in the outer layers. 
As remarkable consequence, all the white dwarf models displayed in Fig.~\ref{fig:omega_comp} have a 
similar amount of total angular momentum 
(see Table~\ref{tab1} and~\ref{tab2}), 
\emph{despite} their different angular momentum distributions.
Fig~\ref{fig:mj1}, where the total angular momentum $J$ of models 
in different model sequences (AAe, ABe, ACe, AFe,  AGe \& Re)
is plotted in the $M - J$ plane,
indeed shows that the total angular momentum, for a given central density,
is a function of the white dwarf mass only, regardless of various possible rotation
profiles. Although the scatter in the value of $J$ becomes larger as the white dwarf mass increases,
it remains within 10 \% from the mean value even when $M \gsim 1.9~\mathrm{M}_\odot$.
Even the rigidly rotating white dwarf models ($M \lsim 1.47$) follow the general trend very well.
We conclude that the $J-M$-law predicted by our models is not sensitive to 
evolutionary details or uncertain physical processes. 

For comparison, we also give results with $j$-constant and $v$-constant rotation laws
in the same $M - J$ plane. 
Note that models with these exotic rotation laws follow very different relations.
However, as mentioned above, these models are shear unstable, and their steep
angular velocity gradients should decay on a short time scale (see Paper~I).
Therefore, we can conclude that $M - J$ relation is not significantly
affected by different $\Omega$-profiles in general. 
We find that this is also the case
for the dependence of other physical quantities on the total mass, 
such as for $T/W$ and and for the binding energy,
as discussed in the following section.

\section{Functional relations of physical quantities}\label{sect:relations}

Here we derive analytical expressions for 
various physical quantities as functions
of the white dwarf mass ($M$)
and the central density ($\rho_\mathrm{c}$), 
from models with the AWD-rotation and rigid rotation laws.

\subsection{$M-J$ relation}
\begin{figure}
\center
\resizebox{\hsize}{!}{\includegraphics{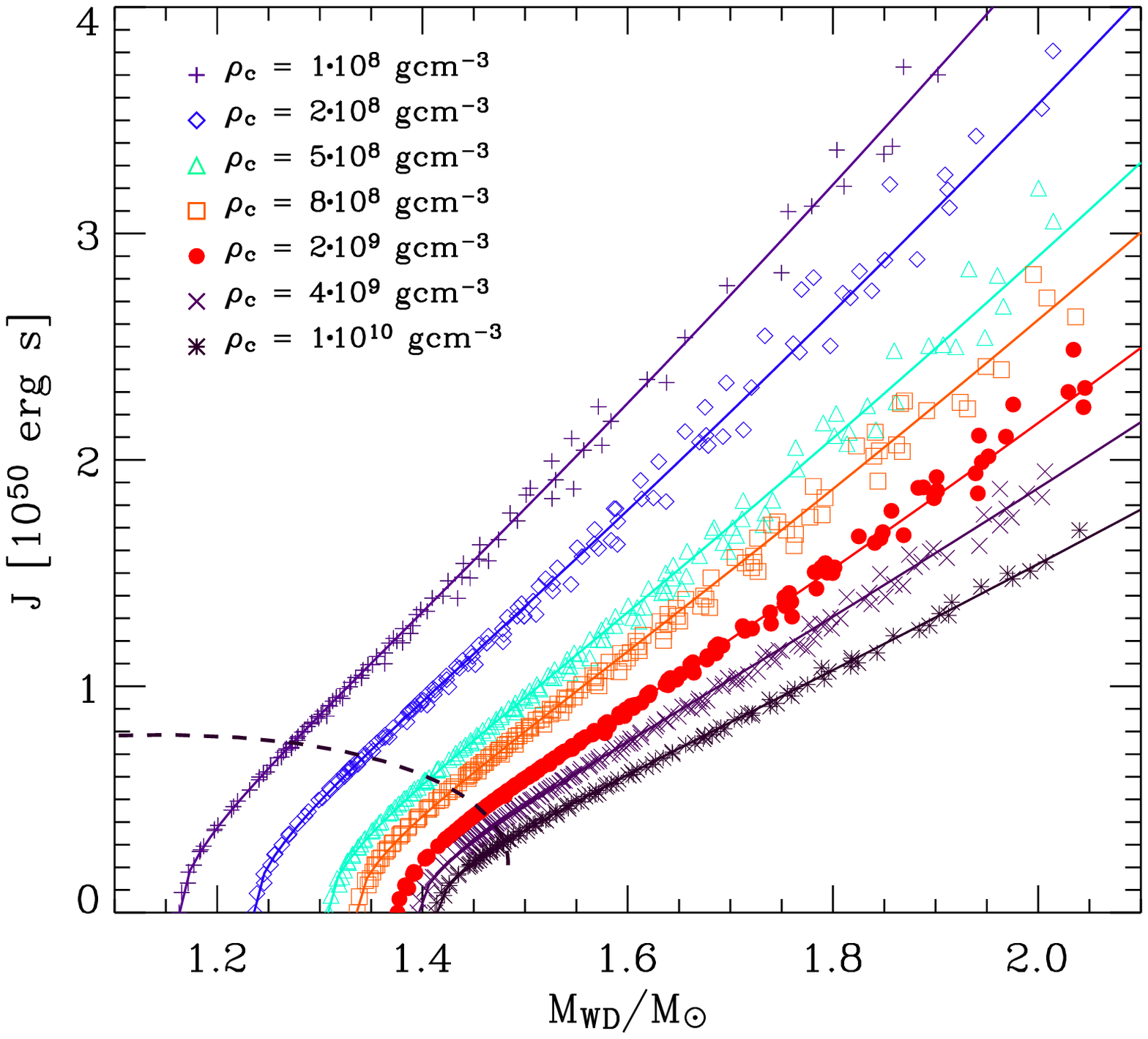}}
\resizebox{\hsize}{!}{\includegraphics{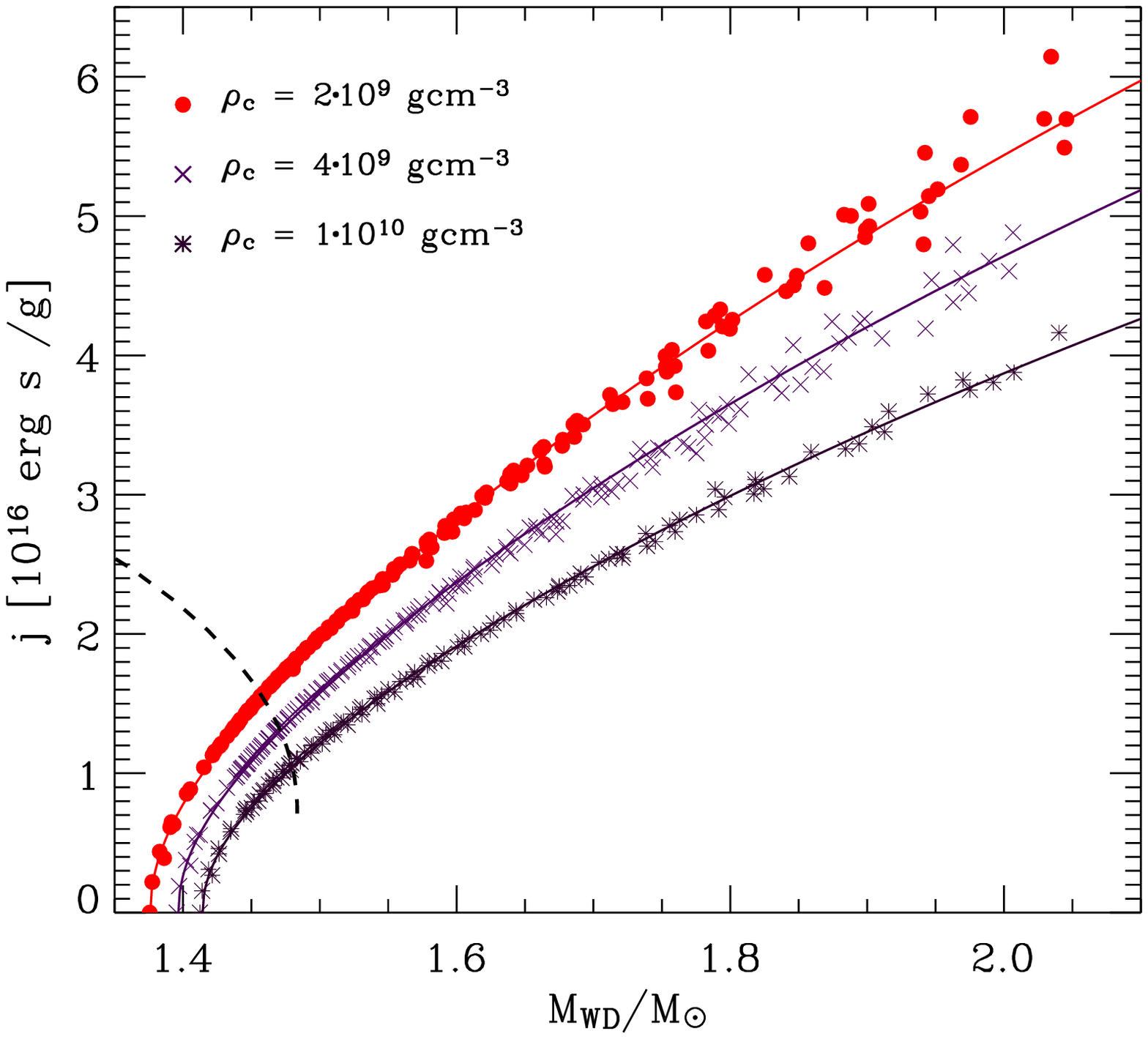}}
\caption{
\emph{Upper Panel:} Total angular momentum of white dwarf models with the AWD rotation
and rigid rotation laws, as a function of the 
white dwarf mass.
Models at different central densities are plotted
with different symbols as indicated in the figure.
The solid lines denote the fitting function given by
Eq.~(\ref{eq19}).
The  dashed line gives the $M-J$  relation
of rigidly rotating white dwarfs at critical rotation.
\emph{Lower Panel:} Specific angular momentum of white dwarf models with the AWD rotation
and rigid rotation laws, as a function of the 
white dwarf mass.
Models with $\rho_\mathrm{c} = 2\times10^9~\mathrm{g~cm^{-3}}$,  
$4\times10^9~\mathrm{g~cm^{-3}}$ 
and $10^{10}~\mathrm{g~cm^{-3}}$ are plotted
with different symbols as indicated in the figure.
The solid lines represent $J(M;\rho_\mathrm{c})/M$, 
where $J(M;\rho_\mathrm{c})$ is given by Eq.~(\ref{eq19}).
The  dashed line gives the $M-j$  relation
of rigidly rotating white dwarfs at critical rotation.
}\label{fig:mj2}
\end{figure}

Fig.~\ref{fig:mj2} shows
the total angular momentum of white dwarf models with the AWD or rigid rotation law
for~7 different central densities, in the $M-J$ plane.
White dwarfs with lower central densities have systematically larger angular momenta
at a given mass, as expected from the role of the centrifugal force.
We find that the relation between the total angular momentum and 
the total mass of the white dwarf 
can be given in an analytical form,  
in the considered range of the central density 
($10^{8} \lsim \rho_\mathrm{c} (\mathrm{g~cm^{-3}}) \lsim 10^{10}$)
and the white dwarf mass ($ M \lsim 2.0~{\mathrm M_\odot}$), as

\begin{eqnarray}\label{eq19}
J (M;\rho_\mathrm{c}) & = &    %
C_1(\rho_\mathrm{c})\left\{1-\exp\left(-0.2\left[M-M_\mathrm{NR}(\rho_\mathrm{c})\right]^{0.48}\right)\right\} %
\nonumber \\
& & + C_2(\rho_\mathrm{c})\left[M-M_\mathrm{NR}(\rho_\mathrm{c})\right]^{1.3}~ \left[10^{50}~\mathrm{erg~s}\right], 
\end{eqnarray}
where 
\begin{equation}
C_1 (\rho_\mathrm{c}) =  20.800370-1.5856256\log_{10}\rho_\mathrm{c}~, 
\end{equation}
and
\begin{equation}
C_2 (\rho_\mathrm{c})  =   11.397626-0.97306637\log_{10}\rho_\mathrm{c}~. 
\end{equation}
Here, $M_\mathrm{NR} (\rho_\mathrm{c})$ denotes the mass of the non-rotating 
white dwarf at a given central density, 
and both $M$ and $M_\mathrm{NR}$ are given in the units of solar masses. 
From non-rotating white dwarf models, 
we derive $M_\mathrm{NR}$ as a function of $\rho_\mathrm{c}$ as
\begin{eqnarray}
M_\mathrm{NR}  &= &  1.436 \times  \nonumber \\
  && \left[1-\exp\left\{-0.01316(\log_{10} \rho_\mathrm{c})^{2.706} 
  +0.2493\log_{10} \rho_\mathrm{c}\right\}\right]   \nonumber \\
 && (\mathrm{for}~\rho_\mathrm{c} > 10^7~\mathrm{g~cm^{-3}})~. 
\end{eqnarray}

\begin{figure}
\center
\resizebox{\hsize}{!}{\includegraphics{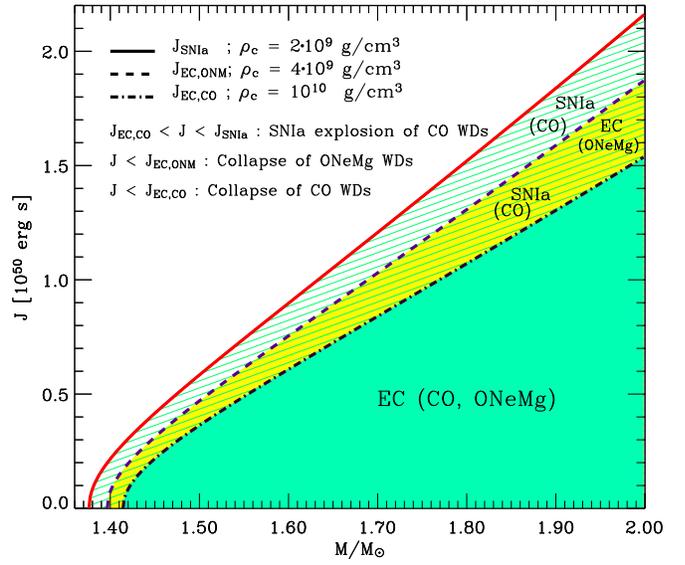}}
\caption{The critical angular momentum for thermonuclear explosion
of CO white dwarfs ($J_\mathrm{SNIa}$; solid line), 
and electron-capture induced collapse
of CO white dwarfs ($J_\mathrm{EC, CO}$; dashed-dotted line) 
and ONeMg white dwarfs ($J_\mathrm{EC, ONM}$; dashed line), 
as a function of the white dwarf mass.
SNIa explosion is expected when $J_\mathrm{EC,CO} \lsim J \lsim J_\mathrm{SNIa}$
(hatched region).
Electron-capture induced collapse is supposed to occur
when $J \lsim J_\mathrm{EC, CO}$ for CO white dwarfs, 
and $J\lsim J_\mathrm{EC, ONeMg}$ for ONeMg white dwarfs, respectively.
}\label{fig:criterion}
\end{figure}

We are particularly interested in the cases of $\rho_\mathrm{c} = 2\times10^9~\mathrm{g~cm^{-3}}$, 
$\rho_\mathrm{c} = 4\times10^9~\mathrm{g~cm^{-3}}$ and $10^{10}~\mathrm{g~cm^{-3}}$, as they represent
critical densities for the thermonuclear explosion of CO white dwarfs, and
the electron-capture induced collapse of ONeMg and CO white dwarfs, respectively
(see Nomoto \& Kondo~\cite{Nomoto91}).
Hereafter, we refer to the total angular momentum at these densities as 
$J_\mathrm{SNIa}$, $J_\mathrm{EC, ONM}$, and  $J_\mathrm{EC, CO}$, respectively:
\begin{equation}
J_\mathrm{SNIa}  = J (M; \rho_\mathrm{c} = 2\times10^9~\mathrm{g~cm^{-3}})~,
\label{eqjsn}
\end{equation}
\begin{equation}
J_\mathrm{EC,ONM} = J (M; \rho_\mathrm{c} = 4\times10^9~\mathrm{g~cm^{-3}})~,
\end{equation}
and
\begin{equation}
J_\mathrm{EC,CO}  = J (M; \rho_\mathrm{c} = 10^{10}~\mathrm{g~cm^{-3}})~.
\end{equation}
If a CO white dwarf has a larger amount of angular momentum than $J_\mathrm{SNIa}$, 
a SN~Ia explosion can not be induced. 
Similarly, an ONeMg white dwarf with $J > J_\mathrm{EC, ONM}$
will not experience electron-capture induced collapse. 
On the other hand, a SN~Ia explosion is expected  when $J_\mathrm{EC, CO} \lsim J \lsim J_\mathrm{SNIa}$.
If carbon ignites at the center when $J < J_\mathrm{EC, CO}$, the outcome will be
electron capture induced collapse.
These criteria for SN~Ia explosion and EC induced collapse
are summarized in Fig.~\ref{fig:criterion}.

Alternatively, the critical CO white dwarf mass for a SN~Ia explosion ($M_\mathrm{SNIa}$)
and ONeMg or CO white dwarf mass for EC induced collapse ($M_\mathrm{EC,ONM}$ or $M_\mathrm{EC, CO}$),  
can be given as a function of $J$, for $M \lsim 2.0~\mathrm{M_\odot}$, as
\begin{equation}
M_\mathrm{SNIa} = 1.376 -0.024J^{0.83}+1-\exp(-0.330J^{1.5})~,
\end{equation}
\begin{equation}
M_\mathrm{EC, ONM} = 1.395 -0.040J^{0.83}+1-\exp(-0.416J^{1.5})~, 
\end{equation}
and
\begin{equation}
M_\mathrm{EC, CO} = 1.412 -0.042J^{0.83}+1-\exp(-0.519J^{1.5})~.
\end{equation}
I.e., exploding/collapsing white dwarfs are expected to have roughly
these critical masses for a given total amount of angular momentum. 

\subsection{$M-T/W$ relation}

\begin{figure}
\center
\resizebox{\hsize}{!}{\includegraphics{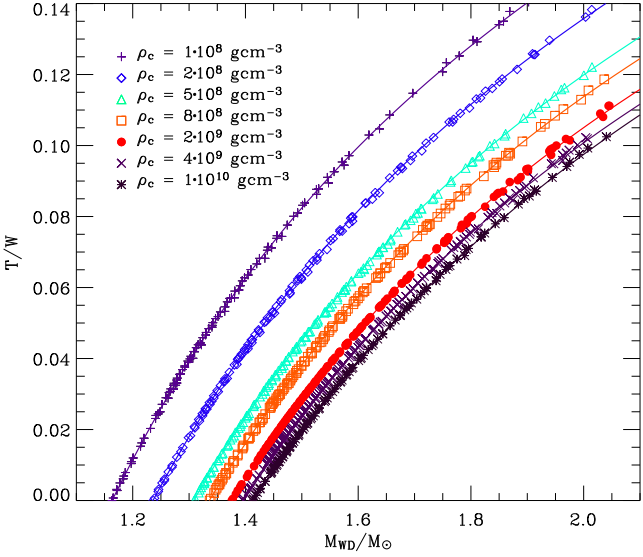}}
\caption{
Ratio of the rotational energy to the gravitational energy ($T/W$) 
in  white dwarf models with the AWD rotation
and rigid rotation laws, as a function of the 
white dwarf mass.
Models at different central densities are plotted
with different symbols as indicated in the figure.
The solid lines denote the fitting function given by
Eq.~(\ref{eq29}).
}\label{fig:tw}
\end{figure}

The ratio of the rotational energy to the gravitational energy ($T/W$)
is plotted as a function of the white dwarf mass in Fig.~\ref{fig:tw}, 
for models with the AWD rotation and rigid rotation laws.
For a fixed central density $\rho_\mathrm{c}$, the values of $T/W$ 
follow very well a one-dimensional relation; the scatter is much smaller 
than in the corresponding diagram showing the total angular momentum
(Fig.~\ref{fig:mj2}). This relation
can be given as a function of $M$ and $\rho_\mathrm{c}$, as
\begin{eqnarray}\label{eq29}
T/W (M;\rho_\mathrm{c})  & = &    %
0.1\left\{1-\exp\left(-C_3(\rho_\mathrm{c})\left[M-M_\mathrm{NR}(\rho_\mathrm{c})\right]\right)\right\} %
\nonumber \\
& & + C_4(\rho_\mathrm{c})\left[M-M_\mathrm{NR}(\rho_\mathrm{c})\right]~, 
\end{eqnarray}
where
\begin{eqnarray}
C_3 (\rho_\mathrm{c}) &=&  407.71741 -168.99280\log_{10}\rho_\mathrm{c} \nonumber \\
           & +& 26.449872(\log_{10}\rho_\mathrm{c})^2  -1.8421546(\log_{10}\rho_\mathrm{c})^3 \\
           & +& 0.048121239(\log_{10}\rho_\mathrm{c})^4 \nonumber
\end{eqnarray}
and
\begin{eqnarray}
C_4 (\rho_\mathrm{c}) &=&  -3.2305060 + 1.5464029\log_{10}\rho_\mathrm{c} \nonumber \\
           & -& 0.26049998(\log_{10}\rho_\mathrm{c})^2  \\
           & + & 0.018811083(\log_{10}\rho_\mathrm{c})^3  - 0.00049405568(\log_{10}\rho_\mathrm{c})^4 \nonumber 
\end{eqnarray}
Note that, in the considered mass range ($M\lsim 2.0 ~\mathrm{M}_\odot$),  
only the models with $\rho_\mathrm{c}=1\times10^8~\mathrm{g~cm^{-3}}$ and 
$2\times10^8~\mathrm{g~cm^{-3}}$ reach the canonical value of 0.14 for the
onset of the bar-mode instability (e.g. Durisen \& Imamura~\cite{Durisen81}). 

\subsection{Binding energy}

\begin{figure}
\center
\resizebox{\hsize}{!}{\includegraphics{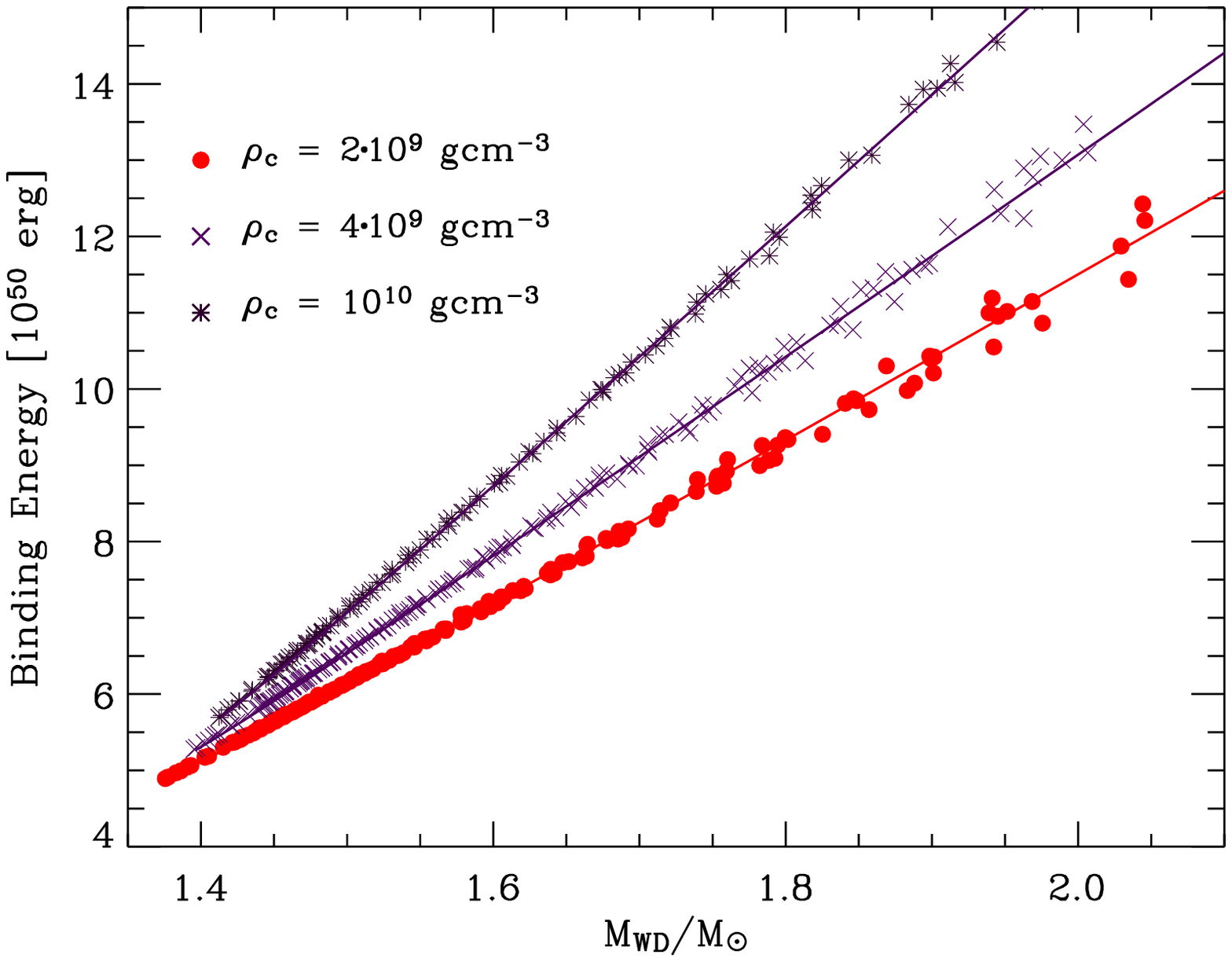}}
\resizebox{\hsize}{!}{\includegraphics{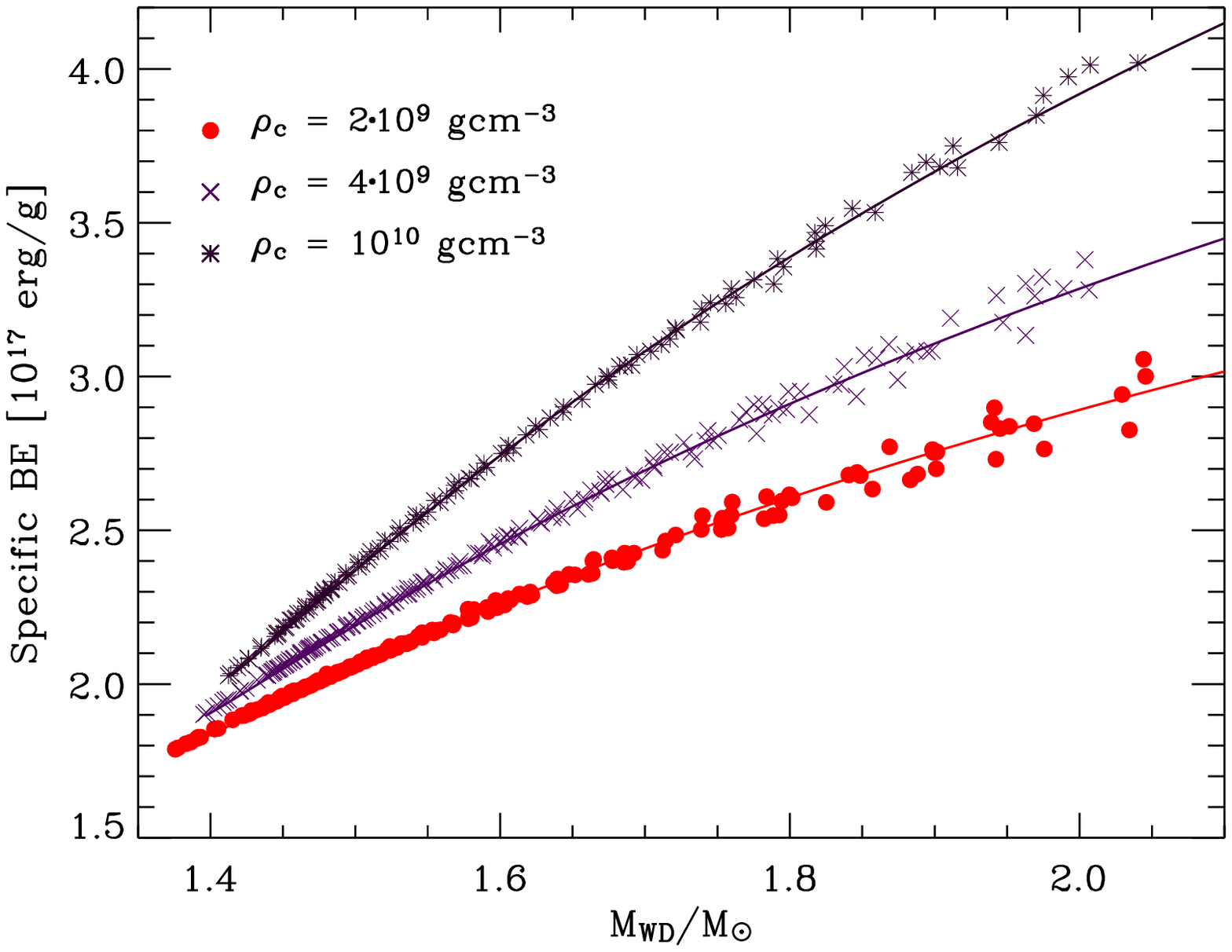}}
\caption{
\emph{Upper panel:} 
Binding energy of white dwarf models with the AWD rotation
and rigid rotation laws for $\rho_\mathrm{c} = 2\times10^9~\mathrm{g~cm^{-3}}$
$4\times10^9~\mathrm{g~cm^{-3}}$, and $10^{10}~\mathrm{g~cm^{-3}}$,
as function of the white dwarf mass.
The solid lines denote the fitting function given by
Eq.~(\ref{eq31}).
\emph{Lower panel:} 
Specific binding energy of white dwarf models with the AWD rotation
and rigid rotation laws for $\rho_\mathrm{c} = 2\times10^9~\mathrm{g~cm^{-3}}$,
$4\times10^9~\mathrm{g~cm^{-3}}$, and $10^{10}~\mathrm{g~cm^{-3}}$.
The solid lines are from the fitting function given by
Eq.~(\ref{eq31}), divided by $M$.
}\label{fig:be}
\end{figure}

The binding energy ($BE := W-U-T$) is essential for the energy budget of exploding or collapsing
white dwarfs. 
Fig.~\ref{fig:be} shows binding energy and specific binding energy 
of our white dwarf models with $\rho_\mathrm{c}=2\times10^9~\mathrm{g~cm^{-3}}$
and $\rho_\mathrm{c}=10^{10}~\mathrm{g~cm^{-3}}$, 
which may represent 
the pre-explosion and pre-collapse stages of CO white dwarfs, respectively, 
and with
$4\times10^9~\mathrm{g~cm^{-3}}$, which refers to 
the pre-collapse stage of ONeMg white dwarfs.
Both, binding energy and specific binding energy
are monotonically increasing functions of the white dwarf mass.

We constructed a fitting formula for the binding energy as a function of
the white dwarf mass, which is applicable in the considered range of central densities 
($1\times10^8 \lsim \rho_\mathrm{c} [\mathrm{g~cm^{-3}}] \lsim 10^{10}$): 
\begin{eqnarray}\label{eq31}
BE(M;\rho_\mathrm{c}) &=& BE_\mathrm{NR}(\rho_\mathrm{c}) \nonumber \\%
 && + C_5(\rho_\mathrm{c})\left[M - M_\mathrm{NR}(\rho_\mathrm{c})\right]^{1.03}~~ \left[10^{50}~\mathrm{erg}\right], 
\end{eqnarray}
where
\begin{eqnarray}
C_5(\rho_\mathrm{c}) &=& -370.73052+132.97204\log_{10}\rho_\mathrm{c} \nonumber \\
                     & - & 16.117031(\log_{10}\rho_\mathrm{c})^2 + 0.66986678(\log_{10}\rho_\mathrm{c})^3 ~.
\end{eqnarray}
The binding energy of non-rotating white dwarfs $BE_\mathrm{NR}$ is given by
\begin{eqnarray}
BE_\mathrm{NR}(\rho_\mathrm{c}) &=& -32.759747+6.7179802\log_{10}\rho_\mathrm{c} \nonumber \\ 
                                & &     -0.28717609(\log_{10}\rho_\mathrm{c})^2~~\left[10^{50}~\mathrm{erg}\right]~. 
\end{eqnarray}

\section{Comparison with 1-D models}\label{sect:comparison}

\begin{figure}
\center
\resizebox{\hsize}{!}{\includegraphics{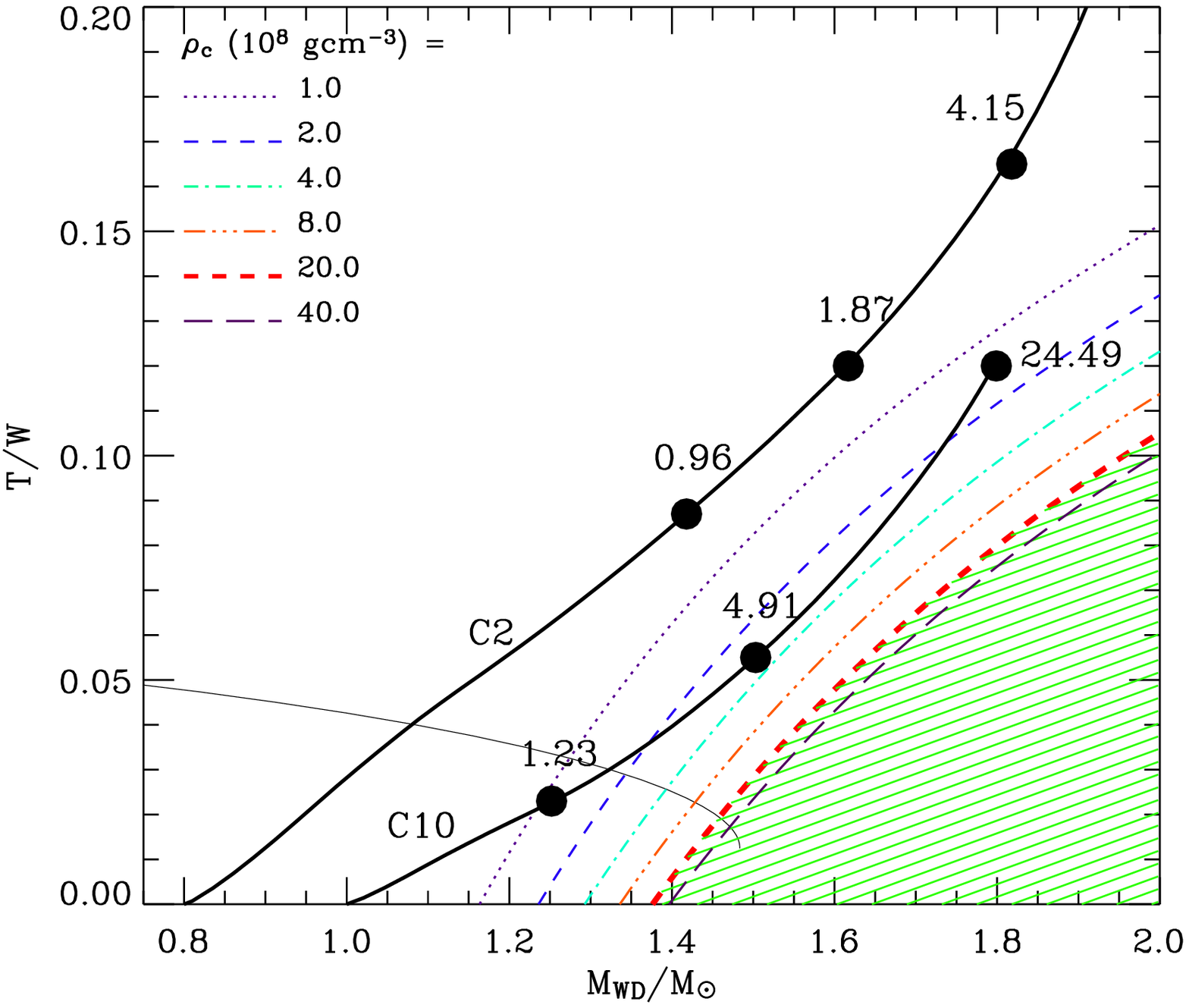}}
\resizebox{\hsize}{!}{\includegraphics{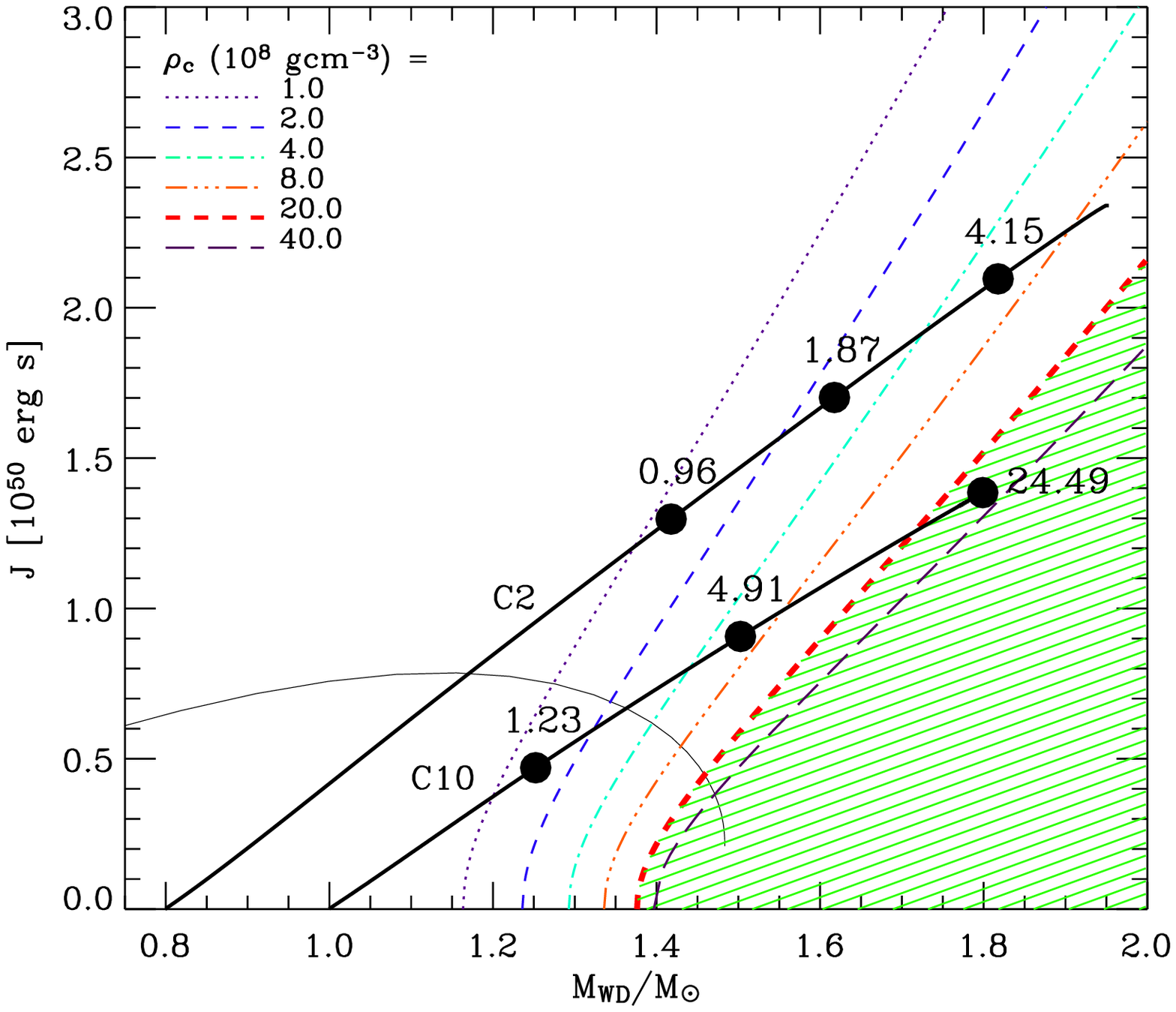}}
\caption{
\emph{Upper panel:} 
Ratio of rotational to gravitational energy ($T/W$) as a function of 
the white dwarf mass. The thin solid line corresponds to the $M-T/W$ relation of 
rigidly rotating 2-D white dwarf models at critical rotation.
Thick solid lines are evolutionary tracks of the 1-D sequences C2 and C10
in Paper~I.
Labels at filled circles on those tracks denote central densities in the unit 
of $10^8~\mathrm{g~cm^{-3}}$.
The $M-T/W$ relation given by Eq.~(\ref{eq29}) is also plotted for different densities:
$\rho_\mathrm{c}/[10^8~\mathrm{g~cm^{-3}}]$ = 1.0 (dotted line), 2.0 (dashed line), 
4.0 (dashed-dotted line), 8.0 (dashed-three-dotted line), 20.0 (thick dashed line)
and 40.0 (long-dashed line). 
\emph{Lower panel:} 
Same as in the upper panel, but in the $M-J$ plane.
}\label{fig:mj4}
\end{figure}

In Fig.~\ref{fig:mj4}, we compare our 2-D results
with with two time-dependent 1-D model sequences of Paper~I in the $M-T/W$ and $M-J$ planes. 
These sequences simulate white dwarfs which accrete with constant accretion rates  
of $5\times10^{-7}~\mathrm{M}_\odot~\mathrm{yr^{-1}}$, 
for an initial white dwarf mass of 0.8~\Msun{} (sequence C2) 
and 1.0~\Msun{} (sequence C10). 
Figure~\ref{fig:mj4} shows that the models of sequence C10 
have similar $T/W$ values at a given mass and a central density
as the 2-D models as long as
$M\lsim 1.6$. However, they deviate significantly for higher masses.
The models of sequence C2, which have a higher 
angular momentum at a given mass compared to those of sequence C10,  
deviate from the 2-D models already when $M\gsim 1.4$. 
We conclude that the 1-D models cannot be regarded as accurate
when $T/W\gsim 0.06$, which occurs after accreting about 0.4~\Msun{} in both cases. 

The inconsistency between 1-D and 2-D models
at $T/W\gsim 0.06$ is not surprising, because
in 1-D models, the effect of the centrifugal force
on the white dwarf structure cannot be accurately described
when a significant fraction of the white dwarf
rotates at above  $\sim$60\% of the Keplerian value (cf. Paper~I).
In particular, the white dwarf radius $R$
is underestimated in this situation. 
In the 1-D models, a certain amount of angular momentum
is assumed to be accreted onto the white dwarf, 
and the consequent angular velocity is determined as 
$\Omega \sim J/(MR^2)$. From 
$T\sim M\Omega^2 R^2 \sim J^2/(MR^2)$ and $W\sim M^2/R$, 
we have $T/W \sim J^2/(M^3R)$.
Therefore, at given $J$ and $M$, an underestimate of $R$ leads to an
overestimate of $T/W$. 

It is remarkable that 
1-D and 2-D models show an extraordinary agreement in the $M-J$ plane
throughout the considered mass range (Fig.~\ref{fig:mj4}, lower panel). 
However, the implication is not that 1-D models 
can describe the white dwarf structure accurately. 
It should be rather considered as coincidence due to the following 
reason. The underestimate of $R$ in 1-D models 
leads to an overestimate of
$\rho_\mathrm{c}$, since it is $\rho_\mathrm{c} \sim M/R^3$. 
On the other hand, the central density should be reduced
due to the overestimate of the centrifugal force, as
the ratio of the centrifugal force to the gravitational force is given
by $F_\mathrm{c}/F_\mathrm{g} \sim \Omega^2R/(GM/R^2) \sim J^2/(M^3R)$.
These two opposite effects might just compensate each other. 

In any case, the consistency between 1-D and 2-D models in the $M-J$ plane
leads us to retain one of the most important conclusions 
of Paper~I:
\emph{If white dwarfs accrete mass and angular momentum efficiently, 
they might hardly reach central carbon ignition 
during the accretion phase.}


\section{Implications for the SNe~Ia progenitor evolution}\label{sect:implication_SNIa}

Our 2-D models of rotating white dwarfs show that 
various physical quantities can be given as a function of 
the white dwarf mass, as long as the shear rate inside the white dwarf
is not stronger than the critical value for the onset of the DSI.
In particular, the $M-J$ relations
of white dwarfs at $\rho_\mathrm{c} = 2\times10^9~\mathrm{g~cm^{-3}}$ 
($J_\mathrm{SNIa}$) and $\rho_\mathrm{c} = 10^{10}~\mathrm{g~cm^{-3}}$ ($J_\mathrm{EC, CO}$)
provides a quantitative criterion for the supernova explosion
of rotating white dwarfs, as summarized in Fig.~\ref{fig:criterion}.
Here, we discuss possible evolutionary paths 
of accreting white dwarfs towards a supernova explosion. 

\subsection{The angular momentum problem}

If a non-rotating white dwarf with the initial mass of $M_\mathrm{init}$ 
accretes mass with a specific angular angular momentum of 
the Keplerian value (i.e., $j_\mathrm{K} = \sqrt{GMR}$), 
the total amount of accreted angular momentum when the white dwarf
reaches mass $M_\mathrm{f}$ is
\begin{eqnarray}
J &=& \int j dM \approx \frac{2}{3}\sqrt{GR}\left[M_\mathrm{f}^{3/2} - M_\mathrm{init}^{3/2}\right] \\
  &\simeq& 4 \times 10^{50} \left(\frac{R}{0.01R_\odot}\right)^{1/2} %
  \left[\left(\frac{M_\mathrm{f}}{\mathrm{M}_\odot}\right)^{3/2} 
       - \left(\frac{M_\mathrm{init}}{\mathrm{M}_\odot}\right)^{3/2} \right]~(\mathrm{erg~s}). \nonumber 
\end{eqnarray}
In binary systems which are considered for the so-called Single Degenerate scenario for SNe~Ia, 
the mass budget is limited, and the maximum possible mass which a CO white dwarf can achieve
by mass accretion is about 2.0~\Msun{} (Langer et al.~\cite{Langer00}). 
Therefore, white dwarfs which are potentially able to perform a SN~Ia explosion
will have masses in the range of $1.4\lsim M/\mathrm{M}_\odot \lsim 2.0$.  
Given that initial masses of $0.8...1.0$~\Msun{} are usually considered in this context
(e.g. Langer et al~\cite{Langer00}), 
the value of $J$ according to the above equation 
will reach $2.6...8.1\times10^{50}~\mathrm{erg~s}$
when the white dwarf mass reaches $1.4...2.0$~\Msun.
This is certainly much larger 
than $J_\mathrm{SNIa}$ (see Fig.~\ref{fig:criterion}), 
and central carbon ignition can not occur in white dwarfs 
with such large angular momenta.

For accreting white dwarfs to end in a SN~Ia,  
therefore, either smaller amounts of angular momentum 
need to be accreted, and/or a large fraction of the accreted 
angular momentum should be lost again.
This means that the detailed evolutionary paths of SNe~Ia progenitors
will depend on the efficiency of the angular momentum gain
and on the time scale of the angular momentum loss.

\subsection{Efficiency of the angular momentum gain}

In Paper~I and also in Uenishi et al. (\cite{Uenishi03}) and Saio \& Nomoto (\cite{Saio04}), 
the outward angular momentum 
transport from the critically rotating star into the accretion disk 
is considered 
to alleviate the angular momentum problem, 
based on the work by Paczy\'nski (\cite{Paczynski91}) and Popham \& Narayan (\cite{Popham91}). 
For instance, in Paper~I, if the equatorial surface of a white dwarf model
rotated at the Keplerian value, angular momentum was assumed to be 
transported outward from the white dwarf
into the accretion disk, while otherwise 
angular momentum of a certain fraction of the Keplerian value was allowed to be accreted.
However, the results of Paper~I showed that even with this limitation, the efficiency  
of the angular momentum gain (hereafter, $\dot{J}_\mathrm{acc}$-efficiency) 
is usually too high for producing SNe~Ia during
the accretion phase:
only when the amount of the accreted angular momentum 
was limited to 30\% of the Keplerian value, 
white dwarfs with a high initial mass ($M_\mathrm{init} = 1.0~\mathrm{M}_\odot$) 
could reach carbon ignition at the center, which occurred at about
$M\simeq 1.8~\mathrm{M}_\odot$ 
(sequences C10, C11 \& C12 in Paper~I; see also Fig.~\ref{fig:mj4}).

It should be noted that 
the angular momentum accretion efficiency might be below 100\%
on two accounts. Firstly, the models of Paper~I neglected thermonuclear shell burning.
Eddington-Sweet circulations will be induced
in the presence of hydrogen and helium shell burning, 
which will accelerate the outward transport of angular momentum
in the white dwarf envelope where the $\Omega$ gradient is negative.
Accordingly, the surface of the white dwarf will rarely deviate from 
rotating critically,
and the angular momentum gain from the accretion disk  will be thus more severely restricted.

Secondly, as pointed out by Saio \& Nomoto (\cite{Saio04}), 
the models of Paper~I applied outward angular momentum transport into the accretion disk 
only for limiting the angular momentum gain when the white dwarf surface reaches 
critical rotation, 
but did not allow for the possibility of removing angular momentum from the white dwarf
into the accretion disk.
Simulations in Saio \& Nomoto (\cite{Saio04}) who did consider the latter possibility
show that the total angular momentum in the accreting white dwarf can not only increase,
but also decrease, depending on the change of the surface conditions.
Systematic studies of the angular momentum accretion efficiency are therefore highly
desirable to improve our understanding of the evolution of SN~Ia progenitors.

\subsection{Gravitational wave radiation}\label{sect:gwr}

As discussed in Paper~I, rotating white dwarfs
can be susceptible to the bar-mode or to the $r$-mode instability, 
which will lead to loss of angular momentum via gravitational wave radiation.
Comparison of Figs.~\ref{fig:mj2} and ~\ref{fig:tw} indicates that
$J\gsim 4 \times 10^{50}~\mathrm{erg~s}$ is necessary for the white dwarf
to reach $T/W = 0.14$, which is the canonical value for the onset of the bar-mode
instability. Although Paper~I described 1-D models with higher values of $T/W$, 
this is due to the overestimate of $T/W$ in the 1-D approximation 
of the stellar structure, as discussed in Sect.~\ref{sect:comparison}. 
In terms of angular momentum, the 1-D models in Paper~I never accrete
more  than $2.5\times10^{50}~\mathrm{erg~s}$, 
implying that $T/W=0.14$ might hardly be achievable.
We conclude that the bar-mode instability may be irrelevant on the SN~Ia context,
and only the $r$-mode instability may be relevant in accreting white dwarfs.

Lindblom (\cite{Lindblom99}) derived the time scale for the growth of $r$-mode instability
in rigidly rotating stars as $1/\tau_\mathrm{r} = K \int\rho(r\Omega)^6 dr$, where
$K = (2\pi/25)(4/3)^8G/c^7$. Using the mean values for density, radius and angular velocity 
($\bar{\rho}$, $\bar{R}$ and $\bar{\Omega}$), this time 
scale can be given as 
$1/\tau_\mathrm{r} \approx K \bar{\rho} \bar{\Omega} \bar{R}^7/7 = (3K/4\pi)M\bar{\Omega}^6\bar{R}^4$. 
Since we have $\bar{\Omega} = J/(kM\bar{R}^2)$, $T = (k/2)M\bar{\Omega}^2\bar{R}^2$
and $W = \gamma GM^2/\bar{R}$, where $k$ and $\gamma$ vary according to the stellar structure,
the time scale $\tau_\mathrm{r}$ can be expressed in terms of $M$, $J$ and $T/W$ as the following:
\begin{eqnarray}\label{eq36}
\log_{10} \tau_\mathrm{r} [\mathrm{yr}] &=& - 3.94 + 10 \log_{10} (J/10^{50}) 
                  - 19\log_{10} (M/\mathrm{M}_\odot)  \nonumber \\
                && - 8 \log (T/W) - 8 \log_{10}\gamma - 2 \log_{10} k~. 
\end{eqnarray}
In our white dwarf models,
$k$ and $\gamma$ are in the range of $0.1 \sim 0.14$ and $1.0\sim 1.6$, respectively.

\begin{figure*}
\center
\resizebox{0.8\hsize}{!}{\includegraphics{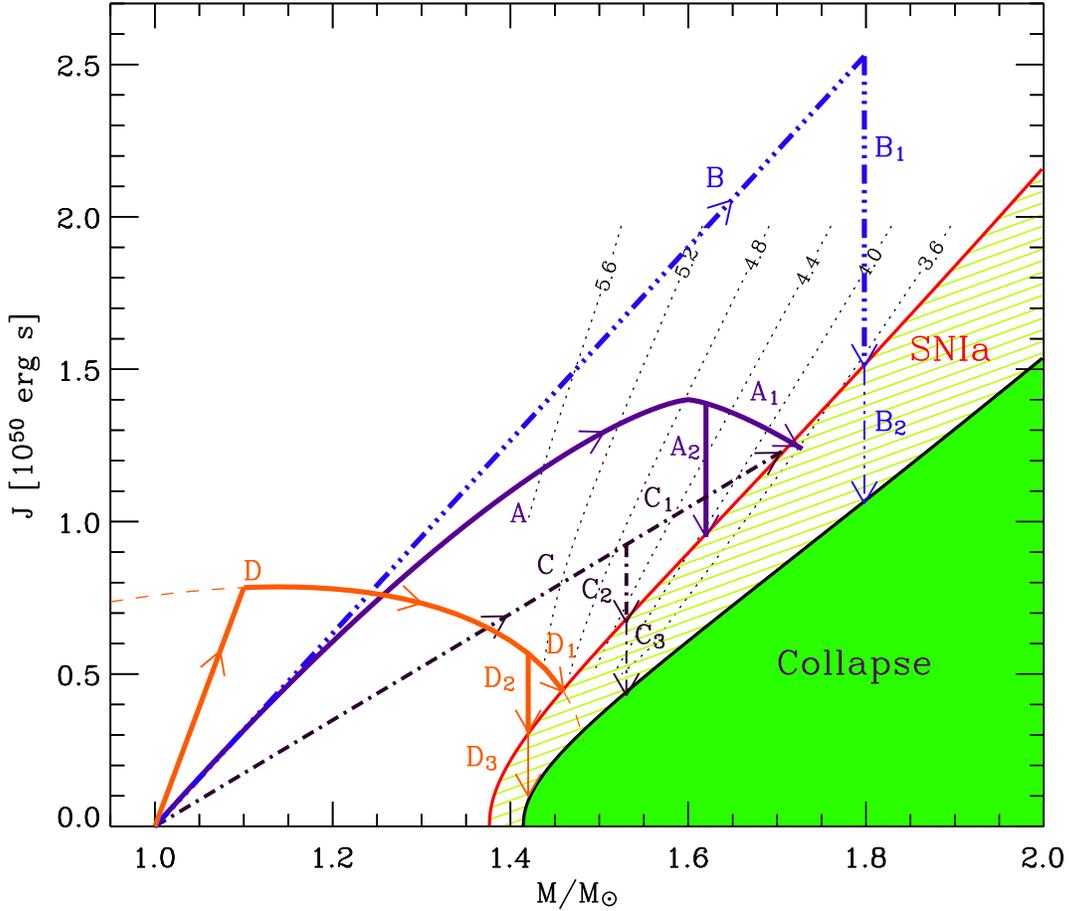}}
\caption{Schematic supernova progenitor evolution scenarios in the $M-J$ plane.
Thick lines starting at a white dwarf mass of 1~\Msun{} with $J=0$
indicate various possible evolutionary paths of accreting white dwarfs:
with angular momentum accretion and simultaneous angular momentum loss (A),
without angular momentum loss during the accretion phase (B), with inefficient
angular momentum accretion (C), and with maximum internal angular momentum transport
efficiency (D); (see text for details).
The region where $J_\mathrm{EC, CO} \lsim J \lsim J_\mathrm{SNIa}$ is hatched,
and the region where $J \lsim J_\mathrm{EC, CO}$ is colored.
For $\dot{M} \simeq 10^{-7}\dots10^{-6}~\mathrm{M~yr^{-1}}$, a supernova explosion
is expected at the upper borderline of the hatched area, $J= J_\mathrm{SNIa}$; cf. Eq.~(\ref{eqjsn}).
The thin dashed line gives the $M-J$ relation of rigidly rotating white dwarfs at critical rotation.
Dotted lines denote lines of constant growth time of the $r$-mode instability ($\tau_\mathrm{r}$; 
Eq.~\ref{eq36}) and are labeled with the logarithm of the growth time in years.  
}\label{fig:taur}
\end{figure*}

In Fig.~\ref{fig:taur}, lines of constant $\tau_\mathrm{r}$ are plotted in the $M-J$ plane, 
in the range of $1.4 < M/\mathrm{M}_\odot < 1.9$ and $0.5 < J/[10^{50}~\mathrm{erg~s}] < 2.0$. 
Here, the $M-J$ and $M-T/W$ relations
given by Eq.~(\ref{eq19}) and~(\ref{eq29}) are used, 
with the fixed values of $\gamma = 1.0$ and $k = 0.14$. 
Values of $\tau_\mathrm{r}$ are found to be about $\sim 10^3 ...  \sim 10^5~\mathrm{yr}$
in the considered range of $M$ and $J$. 
This estimate 
indicates that the $r$-mode instability might have enough time to operate 
in white dwarfs during the accretion phase, 
if $\dot{M} = \sim 10^{-7}~\mathrm{M_\odot~yr^{-1}}$, which is typical for the Single Degenerate
scenario for SN~Ia progenitors.

\subsection{Evolutionary scenarios}\label{sect:evol}

Let us define the accretion time scale 
as $\tau_\mathrm{acc}$, and the time scale for the 
loss of angular momentum due to $r$-mode instability as $\tau_\mathrm{J}$, respectively.

\subsubsection{Case A: $\tau_\mathrm{J} < \tau_\mathrm{acc}$}

When a white dwarf accretes sufficient mass and angular momentum, the
$r$-mode instability begins to operate, and as a consequence the white dwarf 
loses angular momentum. Here we assume that, even though the angular momentum
loss rate is small initially, it will increase during the white dwarf spin-up,
to an extent that $\tau_\mathrm{J} < \tau_\mathrm{acc}$, i.e, that the 
loss of angular momentum occurs more rapidly than its accretion.
Such a scenario is motivated by the r-mode angular momentum loss time scale
as given by Eq.~\ref{eq36}.

In this case, the growth of the white dwarf angular momentum --- and thereby the
growth of its mass --- is limited by the angular momentum loss; it 
ignites carbon at the center when $J$ approaches $J_\mathrm{SNIa}$, 
as illustrated by Path A1 in Fig.~\ref{fig:taur}.
The ending point of Path A1, which defines the explosion mass  $M_\mathrm{A1}$, 
depends on both, the angular momentum loss time scale $\tau_\mathrm{r}$,
and the angular momentum accretion efficiency.
One dimensional evolutionary calculations with
an angular momentum accretion efficiency as in Paper~I 
show that  $\tau_\mathrm{J} \simeq \tau_\mathrm{r}$
results in $M_\mathrm{A1} \simeq 1.6$~\Msun{}
and 1.5~\Msun{}, for the initial masses of 0.8~\Msun{} and 1.0~\Msun{}
(Knaap, Yoon \& Langer, in preparation).
 
In some binary systems, the mass budget may be limited and the 
mass transfer stops before the white dwarf reaches 
$M_\mathrm{A1}$, but at a mass higher than the classical Chandrasekhar mass. 
Then, the white dwarf evolves vertically downward
in the $M-J$ plane as illustrated by path A2. 
Since $\tau_\mathrm{J}  \lsim 10^{6}~\mathrm{yr}$ in this case,
the central temperature at the moment when the white dwarf reaches $J_\mathrm{SNIa}$
is high enough ($T\simeq10^8~\mathrm{K}$) to ignite carbon at 
$\rho_\mathrm{c} \simeq 2\times10^9~\mathrm{g~cm^{-3}}$ (see
Paper~I).
Therefore, both Path A1 and Path A2 lead to the production of SNe~Ia.
The masses of exploding white dwarfs through Path A 
are in the range of $1.38 \mathrm{M_\odot} \lsim M \lsim M_\mathrm{A1}$.
Electron-capture induced collapse can not occur in Case~A.

\subsubsection{Case B: $\tau_\mathrm{J} > \tau_\mathrm{acc}$}

If $\tau_\mathrm{J} > \tau_\mathrm{acc}$, 
the evolutionary path of the accreting white
dwarf will critically depend on the angular momentum accretion efficiency. 
If it is so high that the white dwarf angular momentum 
always exceeds $J_\mathrm{SNIa}$ during the
accretion phase, the white dwarf mass will continue to grow until the mass
transfer comes to an end. Once the mass transfer stops, 
carbon ignition will be delayed until the white dwarf loses
enough angular momentum (Path B in Fig.~\ref{fig:taur}). 
Its final fate will depend on the angular momentum
loss time scale. If it is larger than the white dwarf cooling time 
(i.e., $\tau_\mathrm{J} \gsim 10^9~\mathrm{yr}$), 
carbon ignition will occur after the white dwarf core has crystallized 
--- i.e., when $J \simeq J_\mathrm{EC, CO}$ --- , 
which leads to electron-capture induced collapse and results in a neutron star 
(Nomoto \& Kondo~\cite{Nomoto91}). 
This case is illustrated by Path B2.

If $\tau_\mathrm{J}$ is short enough 
for carbon to ignite in the liquid state in the white dwarf core, 
a SN~Ia explosion will occur (Path B1). 
The mass of exploding white dwarfs
can vary from 1.38~\Msun{} to 2.0~\Msun{}, depending
on the mass budget in the binary system.
However, in this case, the large angular momentum loss time scale of more than 
$10^6\,$yr leads to an ignition density ($\rho_\mathrm{ign}$)
significantly above $\sim 2 \times 10^{9}~\mathrm{g~cm^{-3}}$. 
As SNe~Ia at such high densities produce highly non-solar neutron-rich iron group element ratios
(Iwamoto et al.~\cite{Iwamoto99}),
this scenario can not be considered as the major
evolutionary path.

\subsubsection{Case C: $\tau_\mathrm{J} > \tau_\mathrm{acc}$ and low $\dot{J}_\mathrm{acc}$-efficiency}

If the angular momentum accretion efficiency is sufficiently low, 
white dwarfs can reach $J_\mathrm{SNIa}$ to produce a SN~Ia 
during the accretion phase
even if $\tau_\mathrm{J} > \tau_\mathrm{acc}$, 
as long as enough mass is supplied by the binary companion, 
as illustrated by Path C1.
If the mass transfer stops when $M \gsim 1.38~\mathrm{M_\odot}$ 
but before reaching $J_\mathrm{SNIa}$, 
carbon ignition will be delayed, and its outcome
will depend on the time scale for angular momentum loss, 
as in the Case~B.
If $\tau_{J}$ is shorter than the cooling time scale, a
SN~Ia explosion is expected soon after $J$ reaches $J_\mathrm{SNIa}$ (Path C2). 
Otherwise, carbon burning will occur only when $J \simeq J_\mathrm{EC,CO}$
and result in electron-capture induced collapse to form a neutron star (Path C3). 
The masses of exploding white dwarfs through Path C
are between 1.38~\Msun{} and the ending point
of Path C1 ($M_\mathrm{C1}$).

\subsubsection{Case D: Extreme J-transport efficiency}\label{sect:caseD}

Although Paper~I showed that non-magnetic accreting white dwarfs 
are expected to rotate differentially, the existence of rigidly
rotating white dwarfs, through extremely efficient transport of angular momentum
by magnetic torques, can not be excluded.
If this is the case, the white dwarf surface
begins to rotate critically soon after
the onset of mass accretion,
as shown by  the previous work (Piersanti et al.~\cite{Piersanti03a};  
Uenishi et al.~\cite{Uenishi03}; Saio \& Nomoto~\cite{Saio04})

The angular momentum of rigidly rotating
white dwarf at critical rotation
has an absolute maximum at about 1.13~\Msun{}, 
and decreases as the mass increases further (see Fig.~\ref{fig:mj4}).
Therefore, once a rigidly rotating white dwarf
reaches critical rotation, it must lose
angular momentum during the accretion when $M \gsim 1.13~\mathrm{M_\odot}$.
The $r$-mode instability could remove
angular momentum, if $\tau_\mathrm{J} < \tau_\mathrm{acc}$.
However, in our rigidly rotating white dwarf models
at critical rotation, the estimated growth time scale for the $r$-modes, $\tau_\mathrm{r}$,
becomes smaller than $10^6~\mathrm{yr}$
only when $M \gsim 1.25$~\Msun.
It is therefore questionable whether the $r$-mode instability
can be an efficient mechanism to remove angular momentum during the evolution
of rigidly rotating white dwarfs.
As mentioned earlier, also the bar-mode instability can not operate
as the high required value of $T/W\simeq 0.14$ can not be reached
in rigidly rotating white dwarfs.
Uenishi et al. (\cite{Uenishi03})
and Saio \& Nomoto (\cite{Saio04}) 
considered the viscous transport of angular momentum
from the white dwarf into the accretion disk, 
so that a rigidly rotating white dwarf with $M\gsim 1.13~\mathrm{M_\odot}$ 
may keep rotating critically  with continuous
accretion of mass, while angular momentum decreases progressively.

Within this picture,
the evolution of rigidly rotating white dwarfs
will follow Path D in Fig.~\ref{fig:taur}.
If the binary companion supplies sufficient mass to reach 
$M\simeq 1.46~\mathrm{M_\odot}$, a SN~Ia explosion 
occurs during the accretion phase (Path D1). 
If mass transfer ceases
before reaching $M\simeq 1.46~\mathrm{M_\odot}$ but 
after $M \gsim 1.38~\mathrm{M_\odot}$, 
carbon ignition at the center will be delayed
until enough angular momentum is lost (Paths D2 and D3), 
with the fate of the white dwarf 
depending on $\tau_{J}$, as in the case B.
Exploding white dwarfs through Path D have
masses in the range
$1.38 \lsim M/\mathrm{M_\odot} \lsim 1.46$
(Uenishi et al., \cite{Uenishi03}).

\subsubsection{Diversity in $M_\mathrm{SNIa}$}

The scenarios outlined above show that the masses of exploding white dwarfs ($M_\mathrm{SNIa}$) 
may not be uniform, but are expected to be diverse, with an explosion mass range 
depending on the evolutionary path realized by nature.
The minimum mass is the canonical value of 1.38~\Msun{}, 
while the maximum possible mass ($M_\mathrm{max}$) remains uncertain;
Paths~B and~D result in the largest ($1.38 \lsim M_\mathrm{SNIa} \lsim 2.0~\mathrm{M_\odot}$) 
and smallest ($1.38 \lsim M_\mathrm{SNIa} \lsim 1.46~\mathrm{M_\odot}$) diversity in mass, respectively.
Paths~A and~C may lead to $M_\mathrm{max} < 2.0~\mathrm{M_\odot}$, with
the exact upper mass limit depending on the angular momentum loss time scale and the angular 
momentum accretion efficiency.

\section{Observable consequences of super-Chandrasekhar mass explosions}\label{sect:diversity}

\begin{figure}
\center
\resizebox{\hsize}{!}{\includegraphics{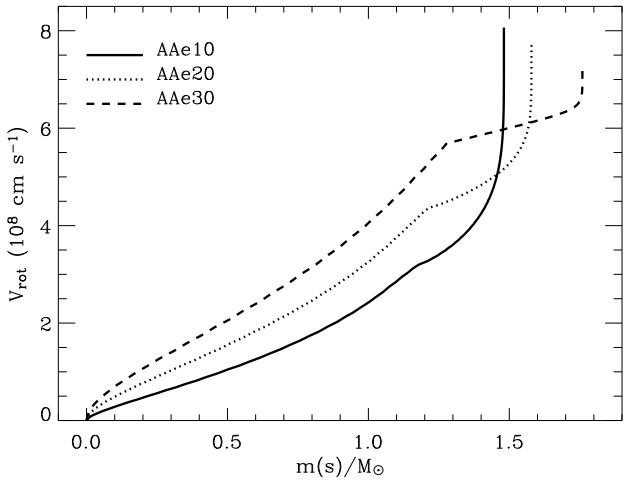}}
\caption{
Equatorial rotational velocity as a function of the mass coordinate
in the pre-supernova models AAe10, AAe20 and AAe30.
}\label{fig:vrot}
\end{figure}

Recent results of 3-D numerical simulations of SN~Ia explosion
indicate that the Delayed Detonation scenario can explain
the major features of observed SNe~Ia, in particular the
explosion energy and nucleosynthesis (Gamezo et al.~\cite{Gamezo04}).
In this scenario, the explosion starts out as a deflagration,
with a deflagration-to-detonation transition (DDT) occurring
during the explosion.
However, so far the DDT does not occur selfconsistently in the
explosion models but needs to be put in by hand.
Lisewski et al.~(\cite{Lisewski00}) showed 
that a DDT in a SN~Ia may occur if turbulent motions with 
velocities of $\sim 10^8~\mathrm{cm~s^{-1}}$ occur on scales of 
$\approx 10^6~\mathrm{cm}$. Such high velocities are not 
expected for Rayleigh-Taylor-driven turbulence in SN Ia deflagration models.
The rotational velocities in our white dwarf models are typically 
from a few to several times $10^8~\mathrm{cm~s^{-1}}$ (Fig.~\ref{fig:vrot}).  
Therefore, even if the white dwarf is spun down by the expansion during the explosion,
rotation could provide enough kinetic energy to enforce the turbulence 
to the level required for a DDT (see also Paper~I).

If the occurrence of a DDT is indeed related to the strength of rotation, 
the DDT might occur earlier (i.e., at higher transition densities) in more massive white dwarfs,  
as those rotate more rapidly (cf. Fig.~\ref{fig:vrot}). 
More massive white dwarfs will produce more 
$^{56}\mathrm{Ni}$ in such a case, 
since 1-D models show that a higher transition density 
results in a more energetic explosion and thus in the production 
of more $^{56}\mathrm{Ni}$ (e.g. H\"oflich \& Khokhlov~\cite{Hoeflich96}).
Even for a pure deflagration, 
more massive white dwarfs are likely to yield more $^{56}\mathrm{Ni}$,  
since they can provide more fuel for the nuclear burning.
Therefore, we may expect the
production of more $^{56}\mathrm{Ni}$,
and thus brighter SNe~Ia (Arnett~\cite{Arnett82}), from more massive white dwarfs.

\begin{figure}
\center
\resizebox{\hsize}{!}{\includegraphics{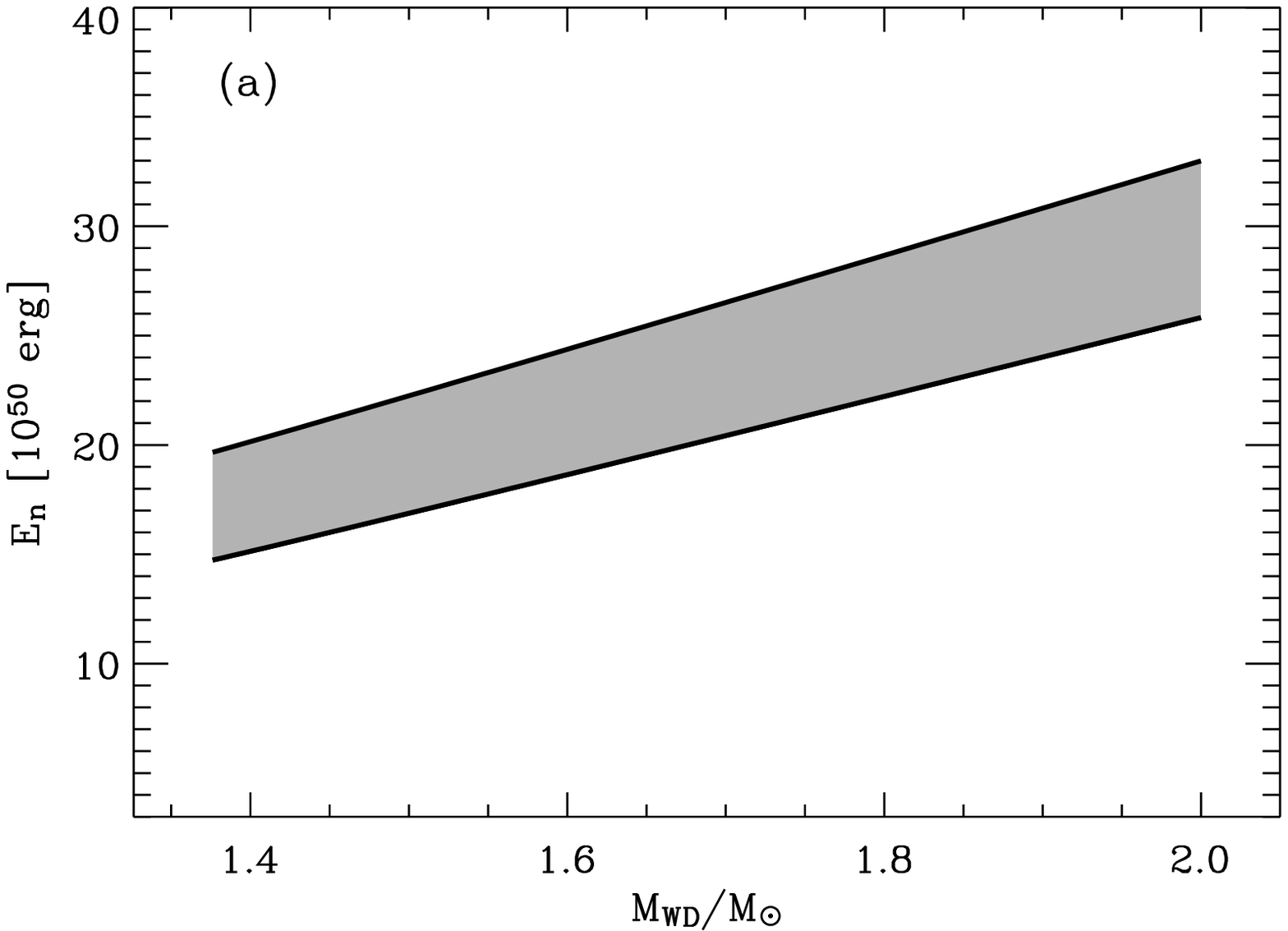}}
\resizebox{\hsize}{!}{\includegraphics{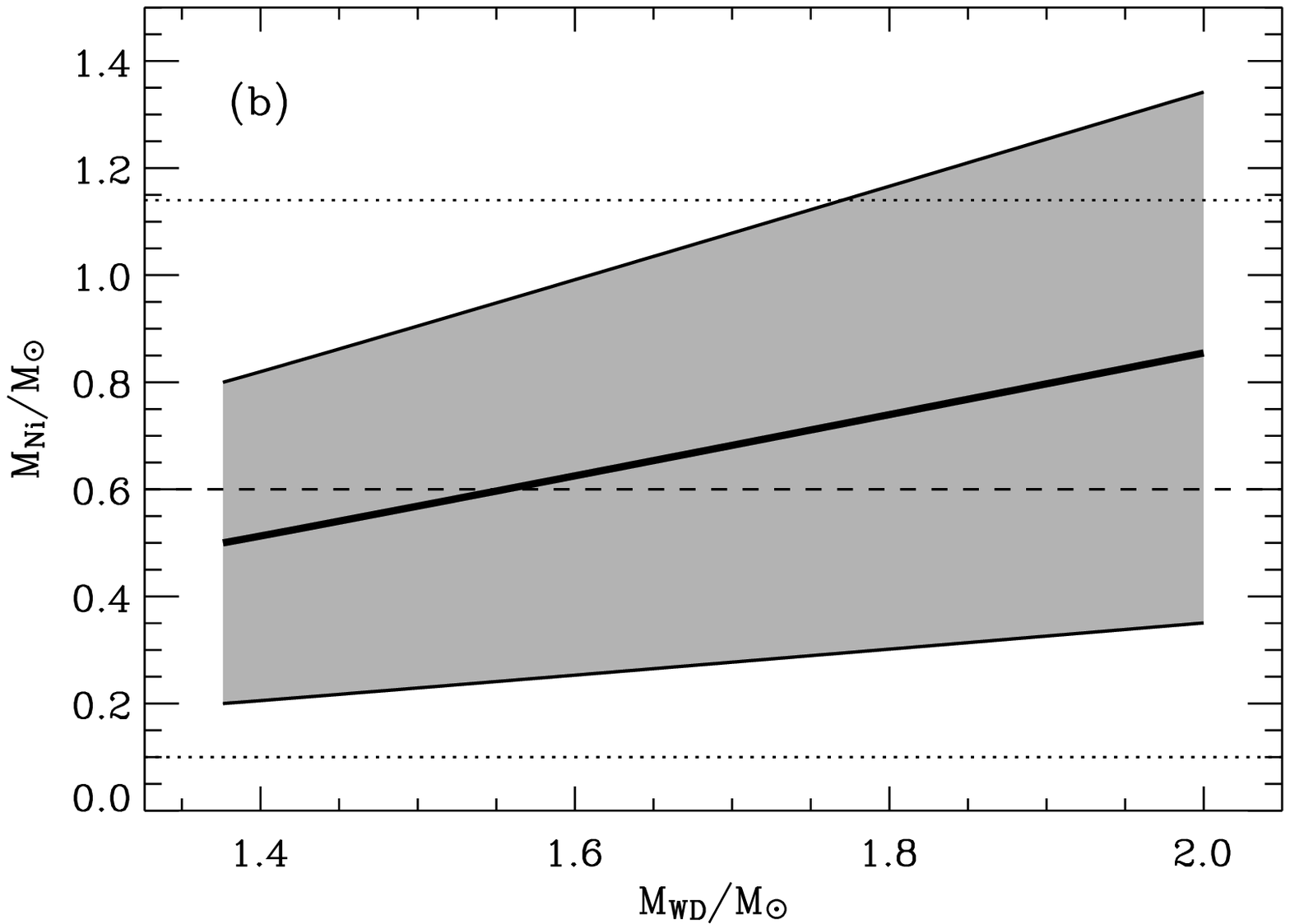}}
\caption{
(a) Nuclear energy required to explain the specific kinetic energy ($e_\mathrm{kin}$)
of observed SNe~Ia, as a function of the white dwarf mass,
i.e, $E_\mathrm{n} = E_\mathrm{BE} + M_{\mathrm WD} e_\mathrm{kin}$. 
The binding energy given by Eq.~(\ref{eq31}) is used here. 
Lower and upper bounds correspond to 
$e_\mathrm{kin} = 3.6 \times10^{17}~\mathrm{erg~g^{-1}}$ 
and $5.4 \times10^{17}~\mathrm{erg~g^{-1}}$, respectively. 
(b) Expected nickel mass produced in a SN~Ia explosion as a function of the white dwarf mass, 
assuming that $M_\mathrm{Ni}$ is proportional to $E_\mathrm{n}$.
Lower and upper limits are computed from the assumption that 
minimum and maximum values for $M_\mathrm{Ni}$
at $M = 1.38$~\Msun{} are 0.2~\Msun{} and 0.8~\Msun{}, respectively.
The thick solid line in the middle of the gray-colored region
defines the mean value between these two limits.
The dashed line gives the mean value of $M_\mathrm{Ni}$ in
observed SNe~Ia. Lower and upper dotted lines indicate 
the observationally estimated minimum 
and maximum $M_\mathrm{Ni}$, 
respectively (Leibundgut~\cite{Leibundgut00}; see text).
}\label{fig:enucl}
\end{figure}

The production of more $^{56}\mathrm{Ni}$ in more massive white dwarfs
is also required for our super-Chandrasekhar mass scenario
to satisfy the observational 
constraints on the kinetic energy ($E_\mathrm{kin}$) of SN~Ia. 
Observed ejecta velocities imply $E_\mathrm{kin} \approx 10^{51}~\mathrm{erg}$, 
assuming that SNe~Ia come from Chandrasekhar mass white dwarfs
(e.g. Branch~\cite{Branch92}). 
In other words, SNe~Ia have a specific kinetic energy ($e_\mathrm{kin}$) of about 
$3.6\times10^{17}~\mathrm{erg~g^{-1}}$. 
No clear correlation between SN~Ia peak brightness and ejecta velocity seems to exist
(e.g. Hatano et al.~\cite{Hatano00}), which might imply that
SNe~Ia have a similar specific kinetic energy on average.
A diversity in the mass of exploding white dwarfs might thus imply that
more massive white dwarfs yield, in general, more specific nuclear
energy ($e_\mathrm{n}$) in the explosion, 
since the specific binding energy ($e_\mathrm{BE}$) increases
with increasing mass (Fig.~\ref{fig:be}).
Although 1-D explosion models  show 
no strict correlation between $e_\mathrm{n}$ and the
produced mass of $^{56}\mathrm{Ni}$ ($M_\mathrm{Ni}$), 
there is certainly a general tendency that a higher
$e_\mathrm{n}$ corresponds to more $M_\mathrm{Ni}$ (e.g. Branch~\cite{Branch92}). 
Therefore, larger nickel masses are expected, on average,
in more massive white dwarfs
to explain the observed SN~Ia kinetic energies. 

Fig.~\ref{fig:enucl}a 
shows the required nuclear energy $E_\mathrm{n}$
to give $e_\mathrm{kin} = 3.6 \dots 5.4 \times10^{17}~\mathrm{erg~g^{-1}}$
as a function of the white dwarf mass, where Eq.~(\ref{eq31}) is used
for the binding energy of white dwarfs
at $\rho_\mathrm{c} = 2\times10^9~\mathrm{g~cm^{-3}}$. 
Here, the adopted range of $e_\mathrm{kin}$ corresponds to what 
is typically assumed in the literature (e.g. Gamezo et al.~\cite{Gamezo04}). 
The variation of $E_\mathrm{n}$ in this figure remains within 75\% in the considered mass range.
1-D Chandrasekhar mass explosion models by H\"oelfich \& Khokhlov (\cite{Hoeflich96})
show that $e_\mathrm{kin} = 3.6 \dots 5.4 \times10^{17}~\mathrm{erg~g^{-1}}$ can 
be achieved with $M_\mathrm{Ni} \approx 0.2 \dots 0.8$, within 
the Delayed Detonation scenario. Assuming the average $M_\mathrm{Ni}$ is 
proportional to $E_\mathrm{n}$, we can estimate a possible range
of $M_\mathrm{Ni}$ which will yield $e_\mathrm{kin} = 3.6 \dots 5.4 \times10^{17}~\mathrm{erg~g^{-1}}$,    
as in Fig.~\ref{fig:enucl}b.
Although the estimated range of $M_\mathrm{Ni}$ is very uncertain
and needs to be confirmed by detailed explosion models, 
this consideration leads us to conclude that rotation
has the potential to plausibly explain the observed diversity
of SN~Ia peak brightnesses: the estimated 
range of $M_\mathrm{Ni}$ covers most of 
the observed range of $0.1~\mathrm{M_\odot} \lsim M_\mathrm{Ni} 
\lsim 1.14~\mathrm{M_\odot}$, 
where the lower and upper limits refer to SN 1991bg
and SN 1991T, respectively (Leibundgut~\cite{Leibundgut00}).

During the last years, differences in the chemical composition 
-- notably, in the C/O ratio or in the overall metallicity --
of exploding white dwarfs have been invoked to explain the
occurrence of different nickel masses for Chandrasekhar-mass white dwarf explosions
(Umeda et al.~\cite{Umeda99}; H\"oflich et al.~\cite{Hoeflich00}; 
Timmes et al.~\cite{Timmes03}). 
However, recent 3-D simulations by R\"opke \& Hillebrandt (\cite{Roepke04})  
indicate that different C/O ratios have a negligible effect
on $M_\mathrm{Ni}$. Also the metallicity dependence appears not
able to explain the full observed spread (Timmes et al.~\cite{Timmes03}). 

Another possible cause for the supernova brightness diversity
within the Chandrasekhar mass scenario
might be related to differences in the carbon ignition density or in the transition density 
for the DDT. However, for any evolutionary path
toward the Chandrasekhar limit, the accretion rates are restricted to
$10^{-7}\dots10^{-6}\mathrm{M_\odot~yr^{-1}}$, since otherwise a white dwarf
can not efficiently grow in mass. 
Since the thermal evolution of non-rotation accreting white dwarfs is mainly determined
by the accretion rate, one can not expect a
significant diversity in the ignition conditions in Chandrasekhar mass white dwarfs. 
Furthermore, 
none of these possibilities could explain the large amount of 
$M_\mathrm{Ni}\approx 1.0~\mathrm{M_\odot}$ in SN 1991T (Fisher et al.~\cite{Fisher99}). 
This leaves a diversity of masses above the classical Chandrasekhar mass,
caused by rotation, as a promising explanation for the origin of the diversity in the SN~Ia brightness. 

Also the diversity in the ejecta velocity 
of SNe~Ia, which was attributed to different explosion mechanisms by
Hatano et al. (\cite{Hatano00}), 
can be related to different white dwarf progenitor masses.
Given that $E_\mathrm{kin} = E_\mathrm{BE} - E_\mathrm{n}$, 
SNe~Ia with the same brightness may 
have somewhat different amounts of kinetic energy, 
if two different masses result in similar
nuclear energy and $M_\mathrm{Ni}$. 
Similarly, it is possible that some
very bright SNe~Ia  have a smaller kinetic energy
than less bright ones, depending on how much nuclear
energy is released at different masses.
Therefore, in our scenario, the 
kinetic energy in SNe~Ia may not necessarily 
correlate with the brightness, 
in accordance with the observations.


The possible correlation between progenitor mass $M_\mathrm{SNIa}$ and the
SNe~Ia brightness ($M_\mathrm{SNIa}-L_\mathrm{SNIa}$ correlation) 
could be tested in various ways. 
Observations indicate that more than 70\% of SNe~Ia are
fairly homogeneous, giving similar brightnesses 
($M_\mathrm{Ni} \simeq 0.6~\mathrm{M_\odot}$) and light-curves, 
while about 20\%...30\% 
are SN 199bg-type (sub-luminous) or SN 1991T-type (over-luminous)
(Hillebrandt \& Niemeyer.~\cite{Hillebrandt00}; Leibundgut~\cite{Leibundgut00}). 
If the $M_\mathrm{SNIa}-L_\mathrm{SNIa}$ correlation were to explain the observed
peak brightness variation, the distribution of progenitor masses ($P_{M_\mathrm{SNIa}}$)
should not be flat, but peak at a certain mass which is to represent the normal SNe~Ia.
Explosions from lower and higher masses may then 
relate to sub-luminous and over-luminous SNe~Ia, respectively.
Population synthesis studies are needed
to investigate whether this scenario 
may result in a progenitor mass distribution which can be consistent with 
the observed distribution of nickel masses. 

It is more straightforward to relate
the $M_\mathrm{SNIa}-L_\mathrm{SNIa}$ correlation
to the fact that SNe~Ia from early type galaxies are
systematically less luminous than those from late type galaxies
(e.g. Ivanov et al.~\cite{Ivanov00}).
Since SN~Ia progenitors must be, in general, much older
in early type galaxies than in late type ones, 
SN~Ia progenitor systems in early type galaxies
should have a systematically smaller mass budget, 
which results in smaller $M_\mathrm{SNIa}$ on average, 
and thus in less luminous SNe~Ia.
Similarly, if the progenitor mass distribution changes systematically
as a function of the metallicity of the progenitor systems, 
SNe~Ia luminosities will also change systematically according to cosmic age. 
For instance, if the initial mass distribution of white dwarfs does not sensitively depend on metallicities, 
we expect the peak of the progenitor mass distribution at a lower mass for lower metallicity 
(Langer et al.~\cite{Langer00}).
We thus expect dimmer SNe~Ia for lower metallicity, which may have important
consequences for the cosmological interpretation of supernova distances at high redshift. 

The aspherical nature of rapidly rotating white dwarfs might, in principle, result in an
observable  polarization, as found 
in SN 1999by (Howell et al.~\cite{Howell01}) and 
SN 2001el (Wang et al.~\cite{Wang03}).
However, the white dwarf rotation per se may not be able to create an aspherical
shape of the ejecta:
Since white dwarf models with $\rho_\mathrm{c} = 2\times10^9~\mathrm{g~cm^{-3}}$
have a specific angular momentum of $\sim 10^{16}~\mathrm{cm^2~s^{-1}}$ 
(Fig.~\ref{fig:mj2}), the rotational velocity of the ejecta will be
$V_\mathrm{ejecta, rot} \simeq j/R_\mathrm{ejecta} \simeq 0.43 ~[\mathrm{km~s^{-1}}] 
(j/3\cdot10^{16}~\mathrm{cm^2~s^{-1}})/(R_\mathrm{ejecta}/10\mathrm{R_\odot})$, 
which is negligible compared to the observed ejecta velocities
in radial direction ($\sim 10^4~\mathrm{km~s^{-1}}$).
Instead, an aspherical distribution of the nuclear fuel in rotating white dwarfs 
(see Fig.~\ref{fig:contour}) and a consequent aspherical explosion
could be more important in producing non-radial movements in the ejecta.
Future 3-D explosion models are needed to relate such effects to our rotating pre-supernova models.

\section{Electron-capture induced collapse}\label{sect:collapse}

Although the present study focuses on Single Degenerate SNe~Ia progenitors,
our 2-D models may also facilitate the study
of double (CO-CO and CO-ONeMg) white dwarf mergers.  During such mergers,
dynamically unstable mass transfer will result in the formation of a thick accretion disk
around the more massive white dwarf (e.g. Benz et al.~\cite{Benz90}).
While it is uncertain how the mass accretion
from the thick disk onto the white dwarf proceeds,
a rapidly spinning super-Chandrasekhar mass merger product can be expected, as
recently shown by Piersanti et al.~(\cite{Piersanti03b}).
In this study, however, the assumption of rigid rotation again strongly restricts
the possible explosion masses. Furthermore, the estimated ratio of rotational
to gravitational energy of $\sim$0.14 in their 1-D rapidly rotating white dwarf
models exceeds by far the maximum possible value for such configurations
(Ostriker \& Tassoul~\cite{OsTa69}; see also Sect.~\ref{sect:caseD}),
casting doubts on the general validity of their results of the
reduced and self-regulated accretion rates due to rotation.
                                                                                                               
Saio \& Nomoto~(\cite{Saio04}) recently suggested that a
double CO-white dwarf merger leads to accretion at
rates near the Eddington limit ($\sim 10^{-5}~\mathrm{M_\odot~yr^{-1}}$),
which, after off-center carbon ignition, will produce a rapidly spinning
ONeMg white dwarf. 
When the central density in such an ONeMg white dwarf
reaches about $4\times10^9~\mathrm{g~cm^{-3}}$, 
electron capture on $^{24}\mathrm{Mg}$ and $^{20}\mathrm{Ne}$
begins, and the collapse of the star to form 
a neutron star is the expected outcome 
(Nomoto \& Kondo~\cite{Nomoto91}).
Strongly magnetized neutron stars with $B\simeq 10^{15}~\mathrm{G}$
and $P\simeq 1 ~\mathrm{ms}$ produced by such events
are proposed as the central engine of cosmic gamma-ray bursts
(e.g. Dar et al.~\cite{Dar92}; Usov~\cite{Usov92}; Duncan \& Thomson~\cite{Duncan92}; 
Yi \& Blackman~\cite{Yi98}; Ruderman et al.~\cite{Ruderman00}).
While non-rotating ONeMg white dwarfs undergo
electron-capture (EC) induced collapse when their mass reaches
the Chandrasekhar-limit, 
rotating ONeMg white dwarfs may need a super-Chandrasekhar mass for such collapse.
Since a single ONeMg white dwarf may be as massive a 1.35\Msun,
the formation of ONeMg white dwarfs as massive as $\sim 2.7~\mathrm{M_\odot}$ 
by mergers is, in principle, possible.

Notably, as shown in Fig.~\ref{fig:mj2},
the specific angular momentum in pre-collapse white dwarfs 
with $M_\mathrm{EC,ONM} \gsim 1.5~\mathrm{M_\odot}$
is  larger than  $j \approx 1.5 \times 10^{16}~\mathrm{cm^2~s^{-1}}$, which
is the maximum angular momentum of a rigidly rotating neutron star.
Therefore, either a large amount of angular momentum must be lost during
the collapse,  or the resulting neutron stars have to rotate differentially. 
In the latter case, the expected mean spin rate
ranges from a few to several times  $\sim 10^4~\mathrm{s^{-1}}$. 

According to our evolutionary scenarios of accreting CO white dwarfs, 
EC induced collapse is also expected to occur if
carbon ignition at the center of a super-Chandrasekhar mass CO white dwarf
is delayed for a long time (Paths B2, C3 and D3; see Fig.~\ref{fig:taur}).
Fig.~\ref{fig:mj2} shows that the specific angular momentum
in pre-collapse CO white dwarfs ($\rho_\mathrm{c} = 10^{10}~\mathrm{g~cm^{-3}}$), 
ranges from $10^{16}~\mathrm{cm^2~s^{-1}}$ to $4\times10^{16}~\mathrm{cm^2~s^{-1}}$
where $1.5\lsim M/\mathrm{M_\odot} \lsim 2.0$.
This is again sufficient for the formation of differentially rotating neutron
stars of with spin rates of $\sim10^4~\mathrm{s^{-1}}$.

In both cases, such rapidly rotating, newly formed neutron stars
are believed to lose angular momentum
via the $r$-mode instability, magnetic dipole radiation, and/or magnetic torques.
More importantly, if the differential rotation could amplify the magnetic field 
up to $10^{15}~\mathrm{G}$  
in such newly born neutron stars (Ruderman et al.~\cite{Ruderman00}), 
or such fields were formed by a dynamo process right after the collapse (Duncan \& Thomson~\cite{Duncan92}),  
the necessary conditions to form a cosmic GRBs
due to electro-magnetic beaming -- 
i.e., $\Omega \approx 10^4~\mathrm{s^{-1}}$ and $B  \approx 10^{15}~\mathrm{G}$ --
might be satisfied (cf. Wheeler et al.~\cite{Wheeler00}; Zhang \& M\'esz\'aros~\cite{Zhang01}).
Although it is a matter of debate whether this channel could be
a major source for GRBs (e.g. Fryer et al.~\cite{Fryer99}; Middleditch~\cite{Middleditch04}),
its possibility deserves careful investigation.
Our 2-D models might provide an ideal starting point for such a study.

\section{Conclusion}\label{sect:conclusion}

We present two-dimensional models of differentially rotating massive white dwarfs,
with internal rotation laws which are derived from evolutionary models of
accreting white dwarfs (Paper~I).
In these models, the shear rate inside the white dwarf is limited by 
the threshold value for the onset of the dynamical shear instability.
We find that various physical quantities of these 2-D white dwarf models
-- such as angular momentum and binding energy --
can be accurately expressed as analytical functions of the 
white dwarf mass and central density, and do not sensitively depend on
the details of the rotation profile (Sect.~\ref{sect:relations}).
In particular, we construct models with $\rho_\mathrm{c} = 2\times10^9~\mathrm{g~cm^{-3}}$
and $10^{10}~\mathrm{g~cm^{-3}}$, which represent the pre-explosion and 
pre-collapse stages of CO white dwarfs, respectively,
and models with $\rho_\mathrm{c} = 4\times10^9~\mathrm{g~cm^{-3}}$ which correspond to the 
pre-collapse stage of ONeMg white dwarfs.
Based on these results, in particular using the mass-angular momentum relation
of our 2-D pre-supernova models, we present new scenarios 
for the evolution of SNe~Ia progenitors (Sect.~\ref{sect:evol}; Fig.~\ref{fig:taur}), in the frame of
a new paradigm of SN~Ia explosion at a variety of super-Chandrasekhar masses. 
By making use of the binding energies of our pre-supernova white dwarf models, we suggest
that the SN~Ia brightness may be proportional to the exploding white dwarf mass on average,
and that this correlation could explain various aspects of the SNe~Ia diversity (Sect.~\ref{sect:diversity}).
We also briefly discussed the electron-capture induced collapse of rapidly rotating
super-Chandrasekhar mass white dwarf
as a potential candidate for the central engine of a sub-class of cosmic gamma-ray bursts
(Sect.~\ref{sect:collapse}).
In conclusion, by providing 
2-D white dwarf models at the pre-explosion and pre-collapse stages (Sect.~\ref{sect:results}),
together with analytic relations describing their global properties (Sect.~\ref{sect:relations}), 
we hope to facilitate future studies of such objects,
including investigations on the role of the $r$-mode instability and
the mass and angular momentum accretion efficiency for the evolution
of accreting white dwarfs, population synthesis models
of the distribution of white dwarf supernova progenitor masses,
its relation to SNe~Ia diversity,
and most importantly, multi-dimensional simulations of SN~Ia explosion and electron-capture induced collapse
of rotating white dwarfs.

\begin{acknowledgements}
This research has been supported in part by the Netherlands Organization for
Scientific Research (NWO).
\end{acknowledgements}


\end{document}